\documentclass[11pt,a4paper]{article}
\pdfoutput=1

\usepackage{amsmath,amssymb}
\usepackage{graphicx}
\usepackage{rotating}
\usepackage{cancel}
\usepackage{bm}
\usepackage[dvipsnames]{xcolor}
\usepackage{comment}
\usepackage{cite}
\usepackage{psfrag}

\usepackage{caption}
\usepackage{subcaption}
\usepackage{enumerate}
\graphicspath{{img/}}

\newcommand{\abs}[1]{\left\lvert{#1}\right\rvert}
\newcommand{\avg}[1]{\left\langle{#1}\right\rangle}
\newcommand{\diff}{\mathrm{d}}
\newcommand{\half}{\frac{1}{2}}

\begin{document}
\numberwithin{equation}{section}

\title{\vspace*{-1.5cm}{\scriptsize  \mbox{}\hfill FTUAM-19-11; IFT-UAM/CSIC-19-74}\\
\vspace{4cm} 
\Large{\textbf{Very Light Asymmetric Dark Matter
\vspace{0.5cm}}}}

\author{Gonzalo Alonso-\'Alvarez$^{1}$, Julia Gehrlein$^{2,3}$, Joerg Jaeckel$^{1}$ and Sebastian Schenk$^{1}$\\[2ex]
\small{\em $^{1}$Institut f\"ur Theoretische Physik, Universit\"at Heidelberg,} \\
\small{\em Philosophenweg 16, 69120 Heidelberg, Germany}\\[0.5ex]
\small{\em $^{2}$Instituto de F\'isica Te\'orica UAM/CSIC,} \\
\small{\em Calle Nicol\'as Cabrera 13-15, Cantoblanco E-28049 Madrid, Spain}\\[0.5ex]
\small{\em $^{3}$Departamento de F\'isica Te\'orica, Universidad Aut\'onoma de Madrid,} \\
\small{\em Cantoblanco E-28049 Madrid, Spain}
\\[0.8ex]}

\date{}
\maketitle

\begin{abstract}
\noindent
Very light dark matter is usually taken to consist of uncharged bosons such as axion-like particles or dark photons.
Here, we consider the prospect of very light, possibly even sub-eV dark matter carrying a net charge that is (approximately) conserved.
By making use of the Affleck-Dine mechanism for its production, we show that a sizable fraction of the energy density can be stored in the asymmetric component.
We furthermore argue that there exist regions of parameter space where the energy density contained in symmetric particle-antiparticle pairs without net charge can to some degree be depleted by considering couplings to additional fields.
Finally, we make an initial foray into the phenomenology of this scenario by considering the possibility that dark matter is coupled to the visible sector via the Higgs portal.

\end{abstract}

\newpage

\section{Introduction}
\label{sec:introduction}
Very light bosons such as axion(-like) particles or dark photons are increasingly popular dark matter candidates (see, e.g.,~\cite{Essig:2013lka,Marsh:2015xka} for reviews).
Currently, a significant and growing experimental community is set to hunt down these very weakly interacting particles~\cite{Sikivie:1983ip,Horns:2012jf,Budker:2013hfa,Jaeckel:2013sqa,Chung:2016ysi,Graham:2015ifn,Kahn:2016aff,TheMADMAXWorkingGroup:2016hpc,Alesini:2017ifp,Melcon:2018dba,Du:2018uak,Jaeckel:2010ni,Hewett:2012ns,Essig:2013lka,Graham:2015ouw,Irastorza:2018dyq}.
Due to their low mass, production of very light dark matter must be non-thermal, as thermal relics with masses significantly below keV are too warm and disrupt structure formation (see, e.g.,~\cite{Colombi:1995ze,Bode:2000gq}).
The most prominent non-thermal production mechanism is the misalignment mechanism~\cite{Preskill:1982cy,Abbott:1982af,Dine:1982ah,Sikivie:2006ni,Nelson:2011sf,Arias:2012az,Jaeckel:2013uva,AlonsoAlvarez:2019cgw}, but other possibilities, such as the decay of cosmological defects~\cite{Ringwald:2015dsf} or precursor particles~\cite{Co:2017mop,Agrawal:2018vin,Dror:2018pdh,Co:2018lka,Bastero-Gil:2018uel,Long:2019lwl}, and production from inflationary perturbations~\cite{Peebles:1999fz,Graham:2015rva,Nurmi:2015ema,Kainulainen:2016vzv,Bertolami:2016ywc,Cosme:2018nly,Alonso-Alvarez:2018tus,Markkanen:2018gcw,Graham:2018jyp,Guth:2018hsa,Ho:2019ayl,Tenkanen:2019aij,AlonsoAlvarez:2019cgw} may contribute to part or even all of the observed dark matter density.

All the aforementioned scenarios have one thing in common: the final dark matter density does not carry any conserved global charge.
As a matter of fact, light dark matter candidates are most commonly considered to be real scalar or vector bosons, which cannot possibly carry any such charge.

In this paper, we consider an alternative situation where most or at least a sizable fraction of the dark matter is asymmetric in the sense that it carries a non-vanishing (approximately) conserved global charge.
This kind of scenario has of course been widely studied under the name of ``asymmetric dark matter''~\cite{Barr:1990ca,Barr:1991qn,Kaplan:1991ah,Gudnason:2006ug,Gudnason:2006yj,Kitano:2004sv,Kitano:2005ge,Kaplan:2009ag,Davoudiasl:2012uw,Petraki:2013wwa,Zurek:2013wia}.
However, in the standard asymmetric dark matter models the particle mass is usually assumed to be in the GeV range, motivated by the fact that a $\sim 5\,{\rm GeV}$ asymmetric dark matter particle would have a (dark) charge asymmetry equal to the baryon asymmetry of the Universe.
We are interested in much lighter particles\footnote{See~\cite{Aguirre:2015mva,HajiSadeghi:2017zrl} for discussions of lighter asymmetric dark matter candidates, that however, do not discuss the origin of the asymmetry.}, typically below an ${\rm MeV}$ and possibly even sub-eV, which due to their high occupation numbers can be described as coherently oscillating classical fields.
Consequently, the asymmetric net charge density is orders of magnitude larger than the baryon asymmetry (and the two charges may not be related to each other at all).
Indeed, for sufficiently small masses, such light dark matter would most likely be carrying the largest charge asymmetry density in the Universe\footnote{For masses $\ll {\rm meV}$, this charge density would be larger than any one potentially carried by neutrinos (or any other fermions), because the energy density of such a fermion gas would exceed the dark matter density due to Pauli blocking, even if the particles were massless.}.

The dark matter and corresponding charge density need to be created in the early Universe. 
The low mass of the particles precludes any purely thermal or even mildly warm production mechanism.
A suitable mechanism -- in spirit very much coherent with the misalignment mechanism -- is the Affleck-Dine mechanism~\cite{Affleck:1984fy} (cf.~\cite{Dine:2003ax} for a review and~\cite{Bell:2011tn,Cheung:2011if,Zurek:2013wia} for its application to asymmetric dark matter).
In this setup, charge is created during the evolution of a field displaced from its minimum in a (slightly) asymmetric potential.
In Section~\ref{sec:charge_generation}, we argue that through this mechanism a sufficient amount of charge can be generated even for very light particles\footnote{An interesting construction where the role of the Affleck-Dine field is played by a pseudo-Nambu-Goldstone boson was recently suggested in~\cite{Harigaya:2019emn}.}.   

Beyond these principal considerations, our main motivation is phenomenological.
Asymmetric dark matter carrying a conserved charge is much harder to detect than its uncharged counterpart, and it requires different methods to do so.
One reason for this is that interactions of dark bosons carrying a conserved charge with the Standard Model are usually of higher dimension than those of a real scalar (or vector) field.
A real scalar field can always couple linearly to any Standard Model singlet operator, $\phi_{\rm real} \, \mathcal{O}_{\rm SM}$.
Coupling a charged scalar to the same operator requires $\phi^{\dagger}\phi \, \mathcal{O}_{\rm SM}$, which is one mass dimension higher and therefore typically more suppressed in the sense of effective field theory.

More important is, perhaps, the fact that many experimentally relevant processes are forbidden by charge conservation.
This is the case too for the standard, comparatively heavy, asymmetric dark matter.
For example and due to the conserved charge, any indirect detection method based on decay or annihilation of dark matter will be suppressed.
At low masses, the effect is even more dramatic.
Since the signal of elastic scattering off Standard Model particles is too small to be detected, essentially all direct detection experiments looking for sub-eV dark matter particles are effectively based on (coherent) absorption of dark matter particles inside the experiment.
For example, the famous axion haloscope~\cite{Sikivie:1983ip} aims to use a strong magnetic field to convert axions into photons, which are then absorbed by the detector.
This, however, is not possible if charge is to be conserved.

This unsuitability of traditional experimental setups leaves us with three options, all of them challenging.
Firstly, there remains the possibility to search directly for the symmetric dark matter fraction, which carries no charge and effectively consists of particle-antiparticle pairs.
Secondly, we can hope that there exist sufficiently unsuppressed charge violating operators that could be exploited in suitably designed experimental searches.
Finally, we can search for other subtle effects that do not require charge violation or even an absorption, e.g.~variations in fundamental constants related to spatial or temporal changes in the dark matter density.
A few initial attempts for the example of a Higgs portal interaction are discussed in Section~\ref{sec:phenomenology}.
Nevertheless, clearly more work is needed to develop effective search strategies for the challenging case of very light asymmetric dark matter.

\bigskip 

The structure of the remainder of this work is as follows. Below, we briefly introduce our model and discuss some of its main features and model building considerations.
In Section~\ref{sec:symmetric_field}, we review the cosmological dynamics of a (homogeneous) self-interacting scalar field in an FLRW background.
Then, in Section~\ref{sec:charge_generation}, we describe how a net charge density can be generated in coherent oscillations of the classical field.
Section~\ref{sec:depletion} considers possible effects that can reduce the energy density stored in the uncharged component of the field while preserving the charge asymmetry, like the resonant annihilation of the symmetric part into an additional scalar field.
In Section~\ref{sec:phenomenology}, we discuss the phenomenology of our scenario focusing on potentially observable effects induced by very light particles with a large charge asymmetry.
The Higgs portal scenario is studied as an interesting example.
Finally, we conclude in Section~\ref{sec:conclusion}.

\subsection{A simple model for very light asymmetric dark matter}

Let us introduce the model under study before going into more detail about its cosmological dynamics in the following sections.
We consider a complex scalar field theory with an approximate $U(1)$ symmetry to play the role of the dark matter,
\begin{equation}\label{eq:lagrangian}
    \mathcal{L} =  \partial_\mu \phi^\dagger \partial^\mu \phi  -m^2\phi^\dagger\phi - \lambda \left(\phi^\dagger\phi\right)^2 -  \epsilon \left(\phi^4 + \phi^{\dagger\,4} \right) \, .
\end{equation}
The first three terms on the right hand side preserve the U(1) symmetry, which is broken explicitly by the quartic term proportional to $\epsilon$.
Without loss of generality, $\epsilon$ can be chosen to be real and positive.

Explicit $U(1)$ breaking terms are necessary in order to generate some amount of conserved charge in a cosmological background.
The reason is that an inflationary epoch in the early Universe dilutes any preexisting $U(1)$ conserved number density.
Therefore, in order to survive until the present day, the charge needs to be replenished after inflation.
As is shown in Section~\ref{sec:charge_generation}, a conserved charge is automatically generated as long as the breaking terms are active at any point after inflation.
This is known as the Affleck-Dine mechanism~\cite{Affleck:1984fy}, by which the initially homogeneous scalar field at rest acquires a non-vanishing angular momentum due to the asymmetry in the shape of the potential caused by the symmetry breaking terms.

The charge generation is most effective when the quartic terms of the potential are relevant.
However, in this regime also other effects due to the self-interaction of the field may become important.
Indeed, when self-interactions are dominating, the dark matter cosmology is affected in two ways.
First, the energy density of a scalar field governed by a quartic term dilutes like radiation.
Second, fluctuations, e.g.~those unavoidably generated during inflation, can grow and cause a fragmentation of the field~\cite{Kofman:1994rk,Shtanov:1994ce,Kofman:1997yn,Kusenko:1997vp,Enqvist:2000cq,Berges:2002cz,Dine:2003ax}.
We recall both effects in Section~\ref{sec:symmetric_field}.

In principle, all $U(1)$ symmetry breaking operators up to fourth order in the fields could be included along with the one in the Lagrangian~\eqref{eq:lagrangian}.
We however only consider the operator $\epsilon \left(\phi^4+\phi^{\dagger4}\right)$ in this work.
The practical reason for this is that in order for the charge to be conserved at late times, the symmetry breaking has to be switched off at some point during the cosmological evolution.
This is achieved when the field value becomes small and the quartic interaction terms are subdominant to the quadratic ones, which are taken to be completely $U(1)$ preserving.
Other quartic terms like $\phi^3\phi^\dagger$ can be introduced without qualitatively changing any of our conclusions, and are kept out of the discussion in the interest of simplicity.

From a theoretical point of view, the choice of the potential~\eqref{eq:lagrangian} can be motivated by symmetry arguments. 
Indeed, the Lagrangian as presented above features an exact $\mathbb{Z}_4$ symmetry.
It has been argued that such discrete symmetries can be preserved even if the global continuous symmetry is broken by quantum gravity effects~\cite{Krauss:1988zc,Kim:2013bla}.
Similar arguments can allow to suppress the symmetry breaking terms even further, e.g.~a $\mathbb{Z}_6$ symmetry would only allow some particular operators of dimension six or higher.
In this sense, our potential is only one instance of this class of models.

The existence of an exact $\mathbb{Z}_4$ symmetry has additional phenomenological consequences, as it severely restricts the possible interaction terms with other particles, most importantly Standard Model ones.
In this case, only symmetry breaking interactions that involve at least four powers of the dark matter field are allowed.
The lowest dimension interactions with Higgs bosons or photons would then be of the form
\begin{equation}
    \frac{1}{M^2}\phi^4 H^{\dagger}H\quad {\rm or}\quad \frac{1}{M^4}\phi^4 F^2 \, ,
\end{equation}
respectively.
In contrast, $U(1)$ symmetric operators are less suppressed by the energy scale $M$,
\begin{equation}
    \abs{\phi}^2H^{\dagger}H\quad{\rm or}\quad \frac{1}{M^2}\abs{\phi}^2F^2 \, .
\end{equation}
As long as $M$ is large, $U(1)$ conserving operators are expected to lead to stronger effects.
We will investigate the phenomenology associated to the Higgs interaction in Section~\ref{sec:phenomenology}. 

\newpage
\section{Cosmological dynamics of a $U(1)$ symmetric field}
\label{sec:symmetric_field}

Let us start with the equations of motion for the complex scalar field~\eqref{eq:lagrangian} in an expanding background.
It is convenient to decompose the field into the radial and the angular degree of freedom, $\phi = \varphi \exp(i\theta) / \sqrt{2}$.
In these coordinates, the equations of motion for a spatially homogeneous\footnote{The assumption of homogeneity of the field at cosmological scales is well motivated. The situation is very similar to that of the misalignment mechanism, since any sufficiently light field present during inflation is homogenized on scales larger than today's horizon.} field read
\begin{align}\label{eq:eom_polar}
	\ddot{\varphi} + 3 H \dot{\varphi} - \varphi \dot{\theta}^2 + \frac{\partial V}{\partial \varphi} = 0 \, , \\
	\ddot{\theta} + \left( 3 H  + 2 \partial_t \log \varphi \right) \dot{\theta} + \frac{1}{\varphi^2} \frac{\partial V}{\partial \theta} = 0 \, ,
\end{align}
where $H=\dot{a}/a$ is the Hubble parameter.

Even though the main focus of this work is the generation of a global $U(1)$ charge density due to an explicit symmetry breaking, we first want to recall the dynamics of the $U(1)$ symmetric case, i.e.~$\epsilon=0$, corresponding to the potential
\begin{equation}\label{eq:symmetric_potential}
	V(\phi) = m^2 \phi^\dagger\phi + \lambda \left( \phi^\dagger\phi \right)^2 \, .
\end{equation}
Looking at~\eqref{eq:eom_polar}, it is clear that the dynamics of a complex field in the above potential can be described using a massive, self-interacting real scalar $\varphi$ and a massless mode $\theta$.
For simplicity, in this section we will assume trivial dynamics of the angular component, $\dot{\theta}=0$, and focus on the radial mode.
This is justified by the fact that during inflation\footnote{We assume that the inflation scale $H_{\rm I} \gg m$.} any initial $\dot{\theta}\neq 0$ and the corresponding conserved charge is diluted, $\dot{\theta} \propto a^{-3}$, therefore quickly becoming negligible.

\subsection{A self-interacting real scalar field}

Once the dynamics of the angular mode are assumed to be trivial, i.e.~$\theta(t) = \mathrm{const}$, the equation of motion for the radial field simplifies to
\begin{equation}\label{eq:eom_U1}
	\ddot{\varphi} + 3 H \dot{\varphi} + m^2 \varphi + \lambda \varphi^3 = 0 \, ,
\end{equation}
which is exactly that of a massive and self-interacting real scalar (as studied originally in~\cite{Peebles:2000yy}).
The existence of the self-interaction causes significant deviations from the usual cosmological evolution of a free scalar field.
The general solution of \eqref{eq:eom_U1} is complicated, but in a similar way to what is commonly done for the misalignment mechanism~\cite{Preskill:1982cy,Abbott:1982af,Dine:1982ah,Sikivie:2006ni,Nelson:2011sf,Arias:2012az}, we can separate the evolution into distinct regimes where solutions are easily found.

At early times, when the Hubble friction is very large compared to the mass and self-coupling terms, Eq.~\eqref{eq:eom_U1} has a dominant constant solution, i.e.~the field is stuck at a certain value, $\varphi(t) = \varphi_0$.
As the Hubble parameter monotonically decreases with time, there comes a point $t_0$ when the friction is not strong enough and the field starts to oscillate around the minimum of its potential.
The Hubble scale at which this transition happens can be estimated as (cf.~\cite{Alvarez:2017kar})
\begin{equation}
	3 H(t_0) \simeq \abs{\frac{\partial^2 V}{\partial \varphi^2}}^{\half} \, .
\end{equation}
Assuming a radiation dominated universe, this corresponds to
\begin{equation}\label{eq:t0}
    t_0 = \frac{3}{2} \abs{m^2 + 3\lambda\varphi_0^2}^{-\half} \, .
\end{equation}
For the rest of this work and unless stated otherwise, the scale factor in the radiation dominated universe will be normalized to $a(t)=\sqrt{t/t_0}$.

In the equation of motion~\eqref{eq:eom_U1}, there is a competition between the mass and the self-coupling of the field.
In the following, we can distinguish two regimes:
the \emph{quartic-dominated} regime when $\lambda \varphi^2 \gg m^2$, and the \emph{mass-dominated} regime, characterized by $\lambda \varphi^2 \ll m^2$.
Given a large enough initial field value, the evolution is at first dominated by the quartic term.
Later on, the mass-dominated regime is eventually reached once the amplitude of the oscillations of the field is sufficiently damped by the expansion of the universe.
If initially $\lambda \varphi_0^2 \ll m^2$, the quartic-dominated regime is skipped and the quartic terms are irrelevant for the cosmological evolution.
As we are interested in studying the effect of the self-interactions, in the following we assume that $\lambda \varphi_0^2 \gtrsim m^2$.

Let us now be more quantitative.
In the quartic-dominated regime, the equation of motion of $\varphi$ can be written as
\begin{equation}\label{eq:eom_U1_quartic}
    \ddot{\varphi} + 3H\dot{\varphi} + \lambda \varphi^3 = 0 \, .
\end{equation}
It is solved by the Jacobi elliptic cosine function with modulus $k^2=1/2$,
\begin{equation}\label{eq:radial_sol_quartic}
	\varphi(\eta) = \frac{\varphi_0}{a} \, \mathrm{cn} \left( \sqrt{\lambda} \varphi_0 \eta, \frac{1}{\sqrt{2}} \right) \approx \frac{\varphi_0}{a} \cos \left( \frac{3}{4} \sqrt{\lambda} \varphi_0 \eta \right) \, ,
\end{equation}
where we use conformal time $\eta=2t_0(a(t)-1)$ (i.e.~$\eta(t_0)=0$) and impose the initial condition $\varphi(0)=\varphi_0$.

The time $t_1$ when the transition from the quartic- to the mass-dominated regime occurs is given by the condition $\lambda \Phi^2(t_1) \simeq m^2$ (where $\Phi$ denotes the amplitude of the oscillations), yielding
\begin{equation}\label{eq:t1}
	t_1 = \frac{\lambda \varphi_0^2}{m^2} t_0 \, .
\end{equation}
In the mass-dominated regime, the Eq.~\eqref{eq:eom_U1} reduces to the well-known damped harmonic oscillator,
\begin{equation}\label{eq:eom_U1_mass}
	\ddot{\varphi} + 3 H \dot{\varphi} + m^2 \varphi = 0 \, ,
\end{equation}
which is solved to an excellent approximation by
\begin{equation}\label{eq:radial_sol_mass}
	\varphi(t) = \varphi_1\left(\frac{a_1}{a(t)}\right)^{3/2} \cos \left( m(t-t_1) \right) \, .
\end{equation}
Here, $\varphi_1$ and $a_1$ are evaluated at time $t_1$.

From the evolution of the field in both regimes, we can obtain the scale factor dependence of the energy density of the field.
In the quartic-dominated regime and averaging over oscillations, the scalar field redshifts as radiation,  $\avg{\rho} \propto a^{-4}$.
In the mass-dominated regime, the energy density dilutes as $\avg{\rho} \propto a^{-3}$, making a massive scalar field a good dark matter candidate\footnote{Note that both results can also be obtained easily by using the virial theorem, instead of explicitly solving the equations of motion.}.

\begin{figure}[t]
\centering
	\includegraphics[width=0.7\textwidth]{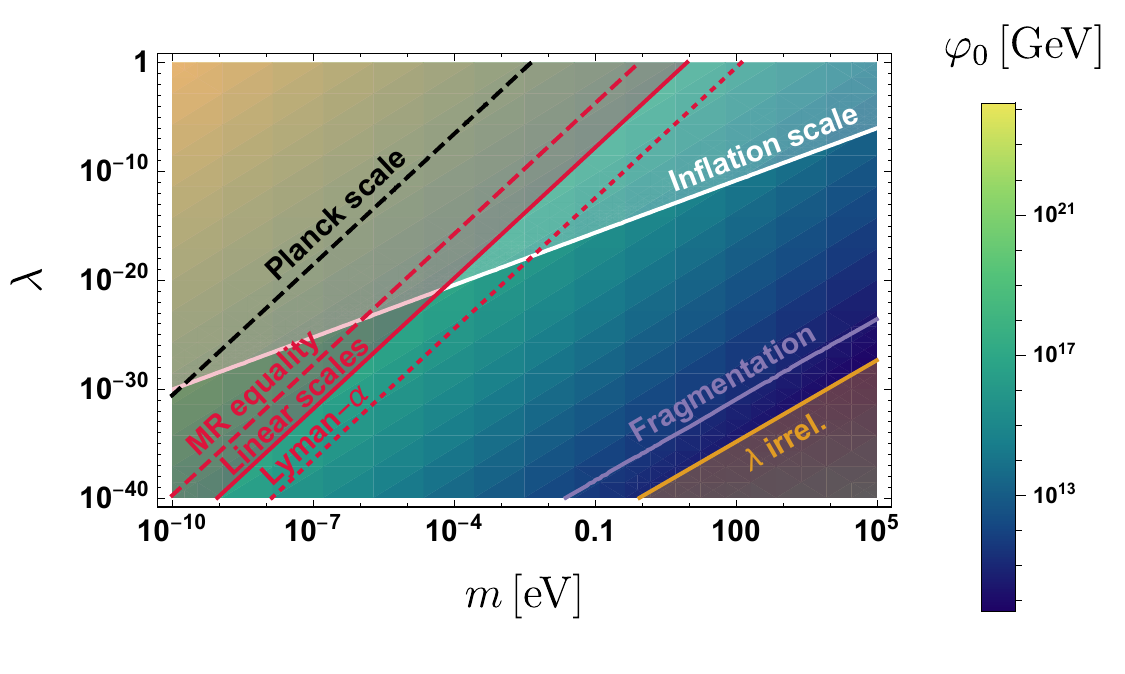}
	\caption{Initial field value $\varphi_0$ of the real scalar field that yields the observed relic abundance of dark matter, shown as a function of the mass $m$ and the self-coupling $\lambda$. The black dashed line illustrates the Planck scale, while the white-shaded region is excluded, because the energy density in the complex field during inflation exceeds that of the inflaton (this bound depends on the inflation scale, for this figure we use $H_{\rm I}=10^{4}\,{\rm GeV}$). In the yellow region the field skips the quartic-dominated regime, such that the self-coupling has no cosmological relevance. The red-shaded region is excluded because the field is still behaving like radiation at the onset of structure formation (more details, including an explanation of the different lines, are given in the main text). Above the purple line the homogeneous field fragments into fluctuations (this again mildly depends on the scale of inflation).}
	\label{fig:phiinitial_U1symm}
\end{figure}

The main consequences of the discussed evolution are summarized in Fig.~\ref{fig:phiinitial_U1symm}.
There, we show the initial field value needed to explain the full dark matter density observed in the universe today, $\rho_{\mathrm{DM}} = 1.3 \, \mathrm{keV}/\mathrm{cm}^3$~\cite{Aghanim:2018eyx}, as a function of the mass and the self-coupling.
Outside of the yellow-shaded region, the field goes through a quartic-dominated phase as described above.
This significantly increases the required size of the initial field value\footnote{In principle, the initial field value could be generated dynamically by the inclusion of a suitable coupling to the Ricci scalar~\cite{Dine:1995kz} of the form $\xi R \phi^\dagger\phi$. This can induce a tachyonic mass during inflation, causing the field to evolve towards its effective minimum. However, for this minimum to be located at the necessary initial field value, the smallness of $\lambda$ in our setup requires the coupling $\xi$ to be extremely small unless inflation happens at a very low scale. This in turn also requires an extremely long period of inflation for the field to reach its effective minimum. Alternatively, if the number of e-folds is not too large, the field could attain a suitable initial condition while still evolving towards the effective minimum for somewhat larger values of $\xi$.} to produce the full dark matter density today, up to the point that above the black dashed line, the initial field value would need to be larger than the Planck scale.
A stronger bound on the initial field value can be obtained from requiring that, during inflation, the energy density stored in the complex field is much smaller than that of the inflaton.
This is satisfied as long as
\begin{equation}
    \lambda \varphi_0 \ll 3 M_{\rm Pl}H_I^2 \, ,
\end{equation}
where $H_I$ is the Hubble scale of inflation.
For a choice of $H_I=10^{4}\,\mathrm{GeV}$, we obtain the bound shown in white in Fig.~\ref{fig:phiinitial_U1symm}.

Furthermore, requiring that the scalar field is sufficiently non-relativistic at the onset of structure formation prevents the quartic-dominated regime from lasting too long. The corresponding limit is shown in red in Fig.~\ref{fig:phiinitial_U1symm}.
A robust bound (solid line) was derived in~\cite{Cembranos:2018ulm}, where a detailed analysis of the effects in the CMB power spectrum and the matter power spectrum at linear scales was performed.
Note that this results in a more stringent constraint than the one coming from the requirement that the field be non-relativistic at the time of matter-radiation equality (dashed line).
An even stronger limit could potentially be placed by using Lyman-$\alpha$ forest observations~\cite{Viel:2005qj}.
This, however, would require a dedicated analysis of the non-linear regime which is beyond the scope of this paper.
Instead, here we merely indicate the region where the field is still relativistic at the time when the smallest scale accessible to Lyman-$\alpha$ observations, $k\sim 40\,\mathrm{Mpc}^{-1}$, enters the horizon (dotted line).

Finally, let us briefly comment on the generation of isocurvature fluctuations in this scenario.
The fluctuations in the dark matter density originate from the quantum fluctuations that $\phi$ acquires during inflation.
Being uncorrelated with the inflaton ones, they are subject to isocurvature constraints.
The density contrast power spectrum at CMB scales has an amplitude $\mathcal{P}_\delta\sim (H_I / \varphi)^2$.
For the parameter region where $\lambda\varphi_0^2\gtrsim m^2$, the observational bound $\mathcal{P}_\delta\lesssim 10^{-10}$~\cite{Akrami:2018odb} is not an issue as long as $H_I\lesssim 10^{6}\,\mathrm{GeV}$, but becomes problematic for larger values of the inflation scale.

\subsection{Fragmentation of the coherent field}
\label{subsec:fragmentation}

Any coherently oscillating, self-interacting scalar field is generically prone to experiencing ``fragmentation'' into inhomogeneous fluctuations that carry non-vanishing momentum~\cite{Kofman:1994rk,Shtanov:1994ce,Kofman:1997yn,Kusenko:1997vp,Enqvist:2000cq,Berges:2002cz,Dine:2003ax}.
This effect can be interpreted as explosive (self-)particle production via a parametric resonance~\cite{Kofman:1994rk,Shtanov:1994ce,Kofman:1997yn,Berges:2002cz} (whose arguments we follow).
In order to assess the relevance of this process, we study the evolution of the fluctuations  $\delta \varphi$ in the homogeneous background field $\varphi$.
Working to linear order in perturbations, $\bar{\varphi}(x) = \varphi(t) + \delta \varphi (\vec{x},t)$, the fluctuations obey the mode equation
\begin{equation}\label{eq:eom_phik_general}
	\ddot{\delta \varphi}_k + 3 H \dot{\delta \varphi}_k + \left( \frac{k^2}{a^2} + 3 \lambda \varphi^2 \right) \delta \varphi_k = 0 \, ,
\end{equation}
which is valid in the quartic-dominated regime.
The dependence of Eq.~\eqref{eq:eom_phik_general} on the expansion of the Universe can be removed by using conformal coordinates, $\hat{\varphi}_k = a\, \delta \varphi_k$ and $\diff \eta = a^{-1} \diff t$, in terms of which the mode equation becomes
\begin{equation}\label{eq:frag_mode_eq_conformal}
	\frac{\diff^2\hat{\varphi}_k}{\diff\eta^2} + \left( k^2 + 3 \lambda a^2 \varphi^2 \right) \hat{\varphi}_k = 0 \, .
\end{equation}
Using the expression Eq.~\eqref{eq:radial_sol_quartic} for the background field in the quartic-dominated regime, Eq.~\eqref{eq:frag_mode_eq_conformal} can be transformed into the following Mathieu equation,
\begin{equation}\label{eq:eom_phik_mathieu}
	\frac{\diff^2 \hat{\varphi}_k}{\diff z^2} + \left( A_k - 2q\cos(2z) \right) \hat{\varphi}_k = 0 \, ,
\end{equation}
with $q=4/3$ and
\begin{align}
	z & = \frac{3}{2} \sqrt{\lambda} \varphi_0\, t_0 \left(a(t) - 1 \right) \pm \frac{\pi}{2} \, , \\
	A_k &= \left( \frac{4}{3} \right)^2 \frac{k^2}{\lambda \varphi_0^2} + 2q \, .
\end{align}
A detailed discussion of the solutions to this equation is given in Section~\ref{sec:depletion} (see also~\cite{Kofman:1994rk,Shtanov:1994ce,Kofman:1997yn,Berges:2002cz}).
Here, and in order to avoid repetitions, we just state the main conclusions.
Floquet's theorem implies that the solution to~\eqref{eq:eom_phik_mathieu} contains an exponential factor,
\begin{equation}\label{eq:Floquet_exponent_fragmentation}
	\varphi_k \propto \exp \left(\mu_k z \right) \, .
\end{equation}
For our case with characteristic parameter $q=4/3$, the dominant contribution to the Floquet exponent $\mu_k$ comes from the second resonance band of the (in)stability chart of the Mathieu equation. The largest exponent corresponds to the mode with momentum $k_\star \sim \sqrt{\lambda} \varphi_0$, for which $\mu \approx 0.1$~\cite{McLachlan:1951}.
Note that $\sqrt{\lambda} \varphi_0\gtrsim m$ in the quartic-dominated regime, which means that the modes that experience the most rapid growth are relativistic at the time of production. 
We can express Eq.~\eqref{eq:Floquet_exponent_fragmentation} in terms of physical time $t$ to obtain the fragmentation rate
\begin{equation}
	\Gamma_{\rm frag}(t) \sim \frac{3\mu}{2} \sqrt{\lambda}\, \frac{\varphi_0}{a(t)} \, .
\end{equation}
If the fluctuations grow too large, most of the energy density that was stored in the homogeneous oscillations is transferred to inhomogeneous and relativistic modes.
Given that our treatment later on relies on the assumption of homogeneity of the field, a complete fragmentation of the field is to be avoided in order to retain analytic control of the dynamics.
Indeed, the linear expansion of the perturbations around the background breaks down even before the mode functions become of size comparable to that of the homogeneous field, i.e.~$\avg{\delta\varphi^2}\sim\varphi^2$.
We expect this to happen at approximately
\begin{equation}
    t_{\rm frag} \sim \Gamma_{\rm frag}^{-1} \,\log\left( \frac{\varphi}{H_I} \right)\, .
\end{equation}
The logarithm takes into account the initial conditions for the growth of the perturbations, which are given by inflationary quantum fluctuations, i.e.~$\delta\varphi_k\sim H_I / k^{3/2}$, where $H_I$ is the Hubble scale of inflation.
The dependence on the initial conditions is relatively weak because of the exponential nature of the fragmentation process.

If fragmentation is significant before the quartic-dominated regime ends, that is, if $t_{\rm frag}\lesssim t_1$, the linear approximation as well as the description in terms of essentially homogeneous fields breaks down. 
The range of parameters where this happens lies above the purple line in Fig.~\ref{fig:phiinitial_U1symm}.
The estimates of the charge generation done in Section~\ref{sec:charge_generation} are therefore only directly applicable between the purple and yellow lines in Fig.~\ref{fig:phiinitial_U1symm}.
We note, however, that this does not necessarily mean that the mechanism is not viable outside of this region.
Proceeding into the non-linear regime would require suitable methods such as classical statistical field theory simulations (cf. e.g.~\cite{Felder:2000hq,Berges:2015kfa}), which are beyond the scope of the present paper.
However, the fragmented regime is not necessarily unsuitable for dark matter.
Indeed, we expect that in large regions of the parameter space the field is still sufficiently cold, but exhibits fluctuations on small yet macroscopic length scales, similar to what was found in~\cite{Berges:2019dgr} for an uncharged field.
For overall attractive interactions, interesting phenomena such as the formation~\cite{Kusenko:1997vp,Enqvist:2000cq} (see also~\cite{Dine:2003ax} for an overview) of Q-balls~\cite{Coleman:1985ki} that can also be dark matter~\cite{Kusenko:1997vp} may arise. In the repulsive case, highly non-trivial effects may also occur~\cite{Moore:2015adu}.
A detailed study of these effects for the very light fields of interest to us would be very interesting.

After understanding the basics of the cosmological evolution of the self-interacting real scalar field, in the next sections we shift the focus to the dynamics of a complex scalar field.

\section{Affleck-Dine dynamics and generation of charge}
\label{sec:charge_generation}

In this section, we study the generation of a net $U(1)$ charge during the cosmological evolution of the complex scalar field with explicit $U(1)$ symmetry breaking terms.
The creation of an effective charge density occurs via the Affleck-Dine mechanism, which was originally proposed as a baryogenesis scenario~\cite{Affleck:1984fy}, but has also been used to link baryogenesis and dark matter~\cite{Bell:2011tn,Cheung:2011if,Zurek:2013wia}. 
Therefore and in contrast to the previous section, we now include the symmetry breaking terms in the potential and take the dynamics of the angular degree of freedom into account.
To be concrete, we consider the model from Eq.~\eqref{eq:lagrangian}, i.e.~a massive scalar field with quartic self-interactions of Affleck-Dine type~\cite{Affleck:1984fy},
\begin{equation}\label{eq:potential}
    V(\phi) = m^2\phi^\dagger\phi + \lambda \left(\phi^\dagger\phi\right)^2 +  \epsilon \left(\phi^4 + \phi^{\dagger\,4} \right) \, .
\end{equation}
The coupling $\epsilon$ explicitly breaks the $U(1)$ symmetry\footnote{If we were to take into account other quartic symmetry breaking terms such as $\phi^{\dagger}\phi^3$, the results would be qualitatively similar.}.
The charge density is produced in the early Universe when this term is active, and is conserved at late times when the field value drops below the threshold at which the quartic interaction becomes inefficient and the $U(1)$ symmetry is effectively restored.
Our goal in what follows is to compute the energy density associated with this $U(1)$-protected charge density. 

\subsection{Charge generation}

Let us now turn to the dynamics of the charge generation for the dark matter field via the Affleck-Dine mechanism, which are essentially unaltered from the case of baryogenesis (see, e.g.,~\cite{Dine:2003ax}).
Importantly, however, we directly consider the homogeneous scalar field as our dark matter candidate. 

Similar to the case of a real scalar, it is convenient to decompose the field into the radial and the angular degree of freedom, $\phi = \varphi \exp(i\theta) / \sqrt{2}$.
In these coordinates, the scalar potential reads
\begin{equation}\label{eq:potential_polar}
	V(\varphi,\theta) = \frac{m^2}{2} \varphi^2 + \frac{\lambda}{4} \varphi^4 + \frac{\epsilon}{2} \cos\left(4\theta\right) \varphi^4 \, .
\end{equation}

Explicitly, the equations of motion for the radial and angular components are
\begin{align}\label{eq:eom_polar_explicit}
	\ddot{\varphi} + 3 H \dot{\varphi} - \varphi \dot{\theta}^2 + m^2 \varphi + \lambda \varphi^3 +  2 \epsilon \cos\left(4 \theta \right)  \varphi^3 = 0 \, , \\
	\ddot{\theta} + \left( 3 H  + 2 \partial_t \log \varphi \right) \dot{\theta}  - 2 \epsilon \varphi^2 \sin(4\theta ) = 0 \, .
\end{align}

The overall evolution of the complex field $\phi$ is similar to the real case.
First, the field is stuck at its initial value and phase due to the Hubble friction.
At a later point, it starts to roll down the potential, probing the quartic self-interactions of the potential that now in addition drive the angular motion via the $U(1)$ symmetry breaking terms.
During this epoch, a significant amount of energy density can be transferred to the angular degree of freedom, creating a net charge asymmetry.
As the Universe expands, the amplitude of the oscillations becomes small enough so that the mass term comes to dominate the evolution and the field acts as dark matter.
In this regime, the effective $U(1)$ symmetry of the theory is restored, finally stabilizing the charge density.

The net $U(1)$ charge density is mostly governed by the dynamics of the angular component~$\theta$,
\begin{equation}\label{eq:charge_explicit}
	n = \varphi^2 \dot{\theta} \, .
\end{equation}
Upon the use of the equations of motion, $n$ satisfies the differential equation
\begin{equation}\label{eq:eom_charge}
	\dot{n} + 3Hn + \frac{\partial V}{\partial \theta} = 0 \, .
\end{equation}
From Eq.~\eqref{eq:eom_charge} it is easy to see that the non-conservation of charge density is proportional to the symmetry breaking coupling,
\begin{equation}\label{eq:eom_charge_formal}
	\frac{1}{a^3} \frac{\diff}{\diff t} \left( a^3 n \right) = - \frac{\partial V}{\partial \theta} =  2 \epsilon \varphi^4 \sin\left(4\theta\right) \, .
\end{equation}
Therefore, in the $U(1)$ symmetric case, $\epsilon=0$, the charge density is comovingly conserved.
From Eq.~\eqref{eq:eom_charge_formal} we can get a qualitative understanding of the behavior of $n(t)$.
The symmetry breaking term on the right hand side sources the charge generation. Depending on $\theta$ it can be positive or negative.
As soon as $\theta$ moves significantly, charge will be continuously created and destroyed for as long as the quartic terms are active.
When the mass term starts to dominate, the field behaves like $\varphi\sim a^{-3/2}$ and the right hand side falls off as $\sim a^{-6}$, effectively restoring the $U(1)$ symmetry.
From then on the previously generated charge density is comovingly conserved.

According to our earlier discussion for the real scalar field, a long enough period of time spent in the quartic-dominated regime can lead to fragmentation of the coherent field.
As shown in Fig.~\ref{fig:phiinitial_U1symm}, the quartic-dominated regime has to be rather short in order to prevent this from happening.
This is possible if the quartic couplings are not much larger than $m^2 / \varphi_0^2$.
In this scenario, the generation of the charge has to be sufficiently fast in order to occur before fragmentation becomes sizeable.
This can be achieved in the band between the purple and yellow regions of Fig.~\ref{fig:phiinitial_U1symm}.
But even when fragmentation is important, there is no reason to believe that charge generation is completely inefficient. 
However, our estimations are not applicable in this case, which requires a full non-perturbative treatment along the lines of~\cite{Felder:2000hq,Enqvist:2000cq,Berges:2015kfa,Moore:2015adu}.
Indeed, interesting non-trivial effects such as a formation of Q-balls or other inhomogeneous structures are expected~\cite{Coleman:1985ki,Kusenko:1997vp,Enqvist:2000cq,Moore:2015adu}.
Yet, this lies beyond the scope of the present work.

\begin{figure}[t]
\centering
	\includegraphics[width=0.5\textwidth]{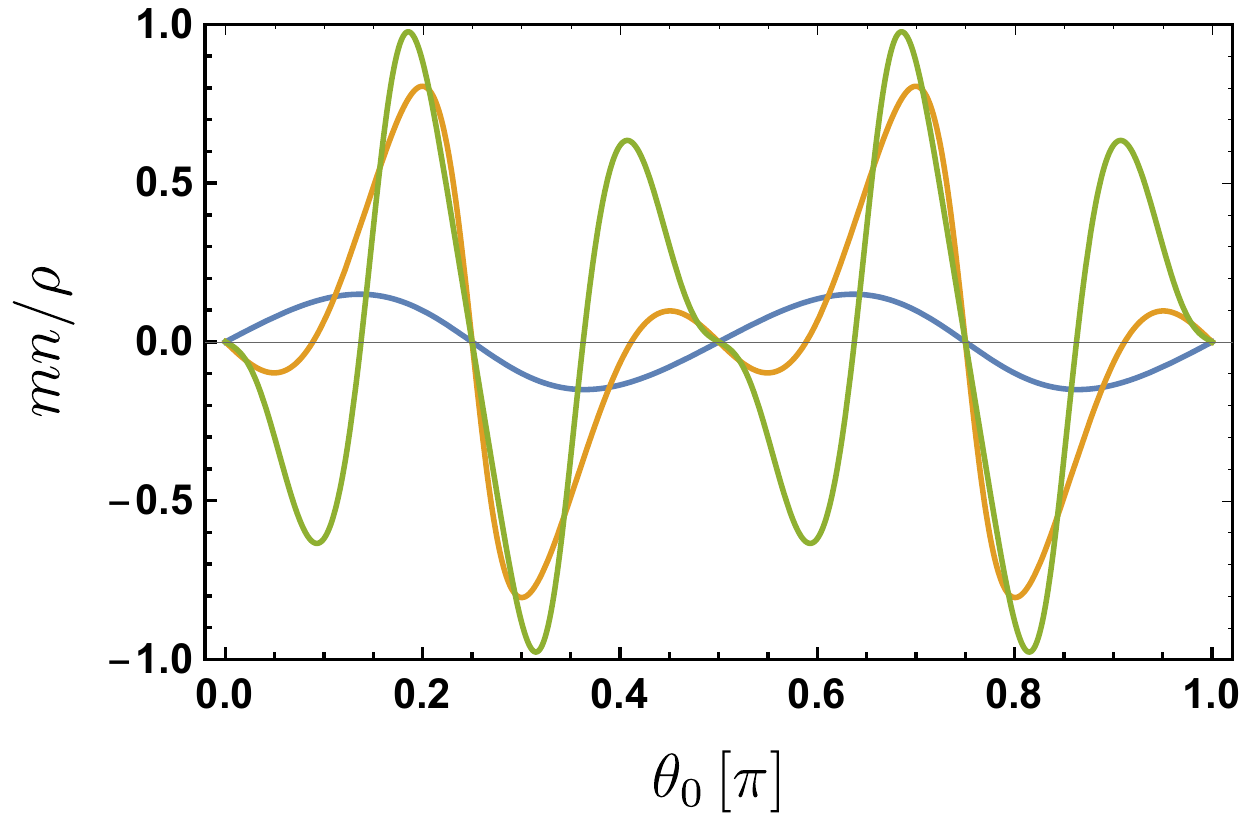}
	\caption{Final fraction of the energy density stored in the charged particles with respect to the total energy density of the coherent oscillations of $\phi$ as a function of the initial misalignment angle $\theta_0$.  The initial radial field value is chosen to give the correct relic abundance of dark matter. As a benchmark, we choose the parameters $m=1$ eV and $\lambda = 10^{-39}$, which fall in the region between the yellow and purple lines in Fig.~\ref{fig:phiinitial_U1symm}. The $U(1)$ breaking coupling $\epsilon$ is set to $0.01 \, \lambda$ (blue), $0.1 \, \lambda$ (yellow) and $0.2 \, \lambda$ (green). As we can see, a sizeable fraction of the energy density can be stored in the charge asymmetry, e.g.~for $\epsilon=0.2\,\lambda$ (green), we observe values in excess of 95\%.}
	\label{fig:nratio_theta}
\end{figure}

\bigskip

In practice, to obtain $n$, we solve the equations of motion for the radial and angular mode numerically and then use Eq.~\eqref{eq:charge_explicit} for the charge density. 
The final amount of net charge is illustrated in Fig.~\ref{fig:nratio_theta}, where we show the fraction of energy density stored in the charge asymmetry, $m\, n$, compared to the total energy density of the field $\rho$, as a function of the initial misalignment angle $\theta_0$ for different values of $\epsilon$.
For simplicity, we estimate the initial field value $\varphi_0$ by considering the real scalar field case illustrated in Fig.~\ref{fig:phiinitial_U1symm}.
For small $\epsilon$ values, the created charge exhibits a simple oscillatory pattern with respect to the initial angle.
The reason is that in this case $\theta$ only traverses a small range of values during the time when the quartic term is active.
Thus, the right hand side of Eq.~\eqref{eq:eom_charge_formal} is always of the same sign and the asymmetry is driven in a single direction.
However, as $\epsilon$ becomes larger, $\theta$ changes more quickly, and leads to a continuous creation and destruction of charge that makes the behavior more complicated.

Remarkably, Fig.~\ref{fig:nratio_theta} shows that it is possible to obtain a sizable fraction close to unity for a suitable choice of parameters without invoking any significant fine tuning.
As already pointed out in the introduction and as will be discussed in Section~\ref{sec:phenomenology}, detecting the part of dark matter that carries a conserved $U(1)$ charge is challenging.
Therefore, if most of the dark matter is charged, current detection techniques may be significantly less sensitive than in purely symmetric scenarios.

\subsection{Charged dark matter}
\label{subsec:charged}

For later use, we now compile some expressions describing the late time behavior of the field.
After the $U(1)$ symmetry is restored, the dynamics are governed by the mass term of the potential.
In this situation, the general solution for the complex field rotates on an ellipse (see, e.g.,~\cite{LandauLifshitz} for useful formulas) in the complex plane,
\begin{equation}\label{eq:ellipse}
    \phi=\left(\frac{a_{1}}{a}\right)^{\frac{3}{2}}\Psi_{1}\left[\cos(\vartheta)\cos(mt+\delta)+i \sin(\vartheta)\cos(mt)\right].
\end{equation}
Here, $\Psi_{1}$ is a complex amplitude and $\delta$ is a phase, both fixed at a suitable time $t_{1}$ or scale factor $a_{1}$.

From Eq.~\eqref{eq:ellipse}, we can easily compute the charge density as
\begin{equation}\label{eq:split}
    n = m\left(\frac{a_{1}}{a}\right)^{3}|\Psi_{1}|^2\sin(2\vartheta)\sin(\delta) \, .
\end{equation}
Similarly, the total energy density is given by
\begin{equation}
    \rho_{\rm CDM}=m^2\left(\frac{a_{1}}{a}\right)^{3}|\Psi_{1}|^2=\rho_{\rm a}+\rho_{\rm s}+{\mathcal{O}}\left(\frac{H}{m}\right) \, ,
\end{equation}
where we neglect small terms arising from the Hubble evolution.
Moreover, we take the opportunity to define the energy densities in the asymmetric and symmetric parts,
\begin{equation}
    \rho_{\rm a}=m n\,,\qquad \rho_{\rm s}=\rho_{\rm CDM}-\rho_{a} \, .
\end{equation}
From this, it is also convenient to introduce the asymmetric and symmetric fractions of the energy density,
\begin{equation}
    \eta_{\rm a}=\frac{\rho_{\rm a}}{\rho_{\rm CDM}}\,,\qquad \eta_{\rm s}=\frac{\rho_{\rm s}}{\rho_{\rm CDM}}\, ,
\end{equation}
which in the quadratic-dominated regime are independent of time.

Later, in Section~\ref{sec:phenomenology} we mostly consider charge conserving interactions.
Those usually involve the combination
\begin{equation}\label{eq:invariant}
    \phi^{\dagger}\phi=\frac{1}{2}\frac{\rho_{\rm CDM}}{m^2}\left[{\mathcal{A}}+{\mathcal{B}}\cos^2(m t+\alpha)\right],
\end{equation}
where,
\begin{equation}\label{eq:ab}
    {\mathcal{A}}=1-\sqrt{1-\eta^2_{\rm a}},\qquad {\mathcal{B}}=2\sqrt{1-\eta^{2}_{\rm a}},
\end{equation}
and $\alpha$ is a phase related to $\vartheta$ and $\delta$, which is however not important for our discussions.

In order to get a better physical understanding of the quantites defined above, let us consider some extreme cases.
The combination in Eq.~\eqref{eq:invariant} is constant if the whole density is stored in the charge asymmetry ($\eta_{\rm a}=1$), whereas in the completely symmetric case ($\eta_{\rm a}=0$) it is purely oscillatory.
More insights can be gained by considering the slightly more general cases $1-\eta_{\rm a}=\eta_{\rm s}\ll 1$ and $\eta_{\rm a}=1-\eta_{\rm s}\ll1$.
In the first case, we have
\begin{equation}
    {\mathcal{A}}\approx 1-\sqrt{2}\sqrt{\eta_{\rm s}\eta_{\rm a}}\,,\qquad {\mathcal{B}}\approx 2\sqrt{2}\sqrt{\eta_{\rm s}\eta_{a}}\,,\qquad{\rm for}\quad (1-\eta_{\rm a})=\eta_{\rm s}\ll 1 \, .
\end{equation}
In other words, there is a large constant part $\sim 1$ while the oscillations are suppressed $\sim\sqrt{\eta_{\rm s}}$. 

In the opposite limit, we have
\begin{equation}
{\mathcal{A}}\approx \frac{\eta_{\rm a}^2}{2}\,,\qquad {\mathcal{B}}\approx 2-\eta^{2}_{\rm a}\,,\qquad{\rm for}\quad \eta_{\rm a}=1-\eta_{\rm s}\ll 1\,.
\end{equation}
This corresponds to oscillations with a small constant component.

\section{Depletion of the uncharged component via annihilations}
\label{sec:depletion}

In contrast to the asymmetric component of the dark matter, the symmetric part is not protected by the $U(1)$ symmetry, so there is the possibility that it depletes, e.g.~via annihilation of the effective\footnote{As long as we are dealing with coherently oscillating fields, one cannot strictly speak of pair annihilation, but it serves as an intuitive picture.} particle-antiparticle pairs.
In this section, we consider such a possibility in order to determine the fraction of the symmetric part of the dark matter that can be lost.
For example, if the annihilation is into massless particles, the energy density stored in them dilutes as radiation and therefore quickly becomes redshifted\footnote{This generally has to happen long enough before matter-radiation equality, but as it was recently shown~\cite{Berezhiani:2015yta}, a small component of dark matter decaying at late times can help alleviate the $H_0$ tension.}.
The result is an effective increase in the fraction of the energy density stored in the charged component compared to the uncharged part.

The specific interaction that we consider in the following does not lead to a full depletion of the uncharged component, as the annihilations stop when the charged and uncharged energy densities are roughly of the same order.
As a consequence, even in situations where the initial charge asymmetry is relatively small, the present day asymmetry can still be significant.
More generally, our discussion provides a proof of principle that a sizable depletion is possible, motivating the thorough investigation of depletion mechanisms.

\subsection{Annihilation into a second (uncharged) scalar}

As a concrete realization of the ideas above, we consider the depletion of the symmetric component by an additional coupling of the complex scalar field $\phi$ to another scalar field $\chi$.
For concreteness, we focus on the interaction
\begin{equation}\label{eq:interaction}
	\mathcal{L}_I = g \abs{\phi}^2 \chi^2 \, .
\end{equation}
In order to study the transfer of energy from the field $\phi$ to $\chi$, we consider the dynamics of the quantum field $\chi$ in the oscillating background $\phi$.
Such a situation has been intensely studied in the context of reheating after inflation~\cite{Kofman:1994rk,Shtanov:1994ce,Kofman:1997yn,Berges:2002cz} (cf.~also~\cite{Baumann}), and we base our discussion on the findings of these works.

The interaction term~\eqref{eq:interaction} induces an oscillating mass term for the field $\chi$, i.e.~the frequencies of the mode functions of $\chi$ are modified as
\begin{equation}
	\ddot{\chi}_k + 3H\dot{\chi}_k + \left( \frac{k^2}{a^2} + m_{\chi}^2 + 2 g \abs{\phi}^2 \right) \chi_k = 0 \, .
\end{equation}
It is then possible that some momentum modes are enhanced by resonance effects.
Since the mode functions determine the occupation number, this enhancement can be viewed as a  resonant production of $\chi$-particles.
This process is known as \emph{parametric resonance}~\cite{Kofman:1994rk,Shtanov:1994ce,Kofman:1997yn,Berges:2002cz}.

Let us therefore study the parametric resonance induced by the coherent oscillations of $\phi$ in more detail.
For simplicity, we take $\chi$ to be massless, such that the mode functions obey
\begin{equation}\label{eq:eom_chik_general}
	\ddot{\chi}_k + 3H\dot{\chi}_k + \left( \frac{k^2}{a^2} + g \varphi^2 \right) \chi_k = 0 \, ,
\end{equation}
where we use the polar decomposition of $\phi$.
From the above equation of motion it is clear that the oscillating nature of $\varphi$ can drive the $\chi$-modes into a resonance, inducing rapid (as we will shortly see, exponential) growth of the mode functions.

The total number density of $\chi$-particles is given by a summation over all modes,
\begin{equation}\label{eq:chi_numberdensity}
	n_\chi = \frac{1}{\left( 2 \pi a \right)^3} \int \mathop{\diff^3 k} n_k = \frac{1}{2\pi^2 a^3} \int \mathop{\diff k} k^2 n_k \, .
\end{equation}
The corresponding energy density of the generated $\chi$-particles can be computed similarly,
\begin{equation}\label{eq:chi_energydensity}
	\rho_\chi =  \frac{1}{\left( 2 \pi a \right)^3} \int \mathop{\diff^3 k} \omega_k n_k = \frac{1}{2\pi^2 a^3} \int \mathop{\diff k} k^2 \omega_k n_k \, ,
\end{equation}
where $\omega_k$ denotes the energy of the mode.
The comoving occupation numbers can be obtained from the mode functions via~\cite{Kofman:1997yn},
\begin{equation}
    n_{k}=\frac{\omega_{k}}{2}\left(\frac{|\dot{\chi}_{k}|^2}{\omega^2_{k}}+|\chi_{k}|^2\right)-\frac{1}{2}.
\end{equation}
Proceeding in a similar way as we did in Section~\ref{sec:symmetric_field}, we study separately the evolution of the occupation numbers governed by~\eqref{eq:eom_chik_general} in the two different regimes of the evolution of the radial mode $\varphi$: the quartic- and mass-dominated regimes.
In the following, for simplicity, we neglect the asymmetric part of the field and only consider the symmetric contribution. 
That is, we focus on the depletion of the symmetric part of $\phi$ (cf.~Section~\ref{subsec:charged}).
We comment on the effects of a non-vanishing asymmetric part at the end of this section.

\subsection{Quartic-dominated epoch}

In the regime where the quartic self-interaction dominates the evolution of the field, the theory is essentially scale invariant.
Hence it is useful to use conformal coordinates $x_k = a \chi_k$ and $\diff \eta = a^{-1} \diff t$, in which the equation of motion reads,
\begin{equation}\label{eq:eom_chik_quarticdominated}
	\frac{\diff^2 x_k}{\diff \eta^2} + \left( k^2 + g a^2 \varphi^2 \right) x_k = 0 \, .
\end{equation}
Importantly, we assume that the evolution of the radial mode is governed by the $U(1)$ symmetric interactions only, i.e.~it is given by~\eqref{eq:radial_sol_quartic}. Under this assumption, the equation of motion for the mode functions~\eqref{eq:eom_chik_quarticdominated} can be rewritten as a Mathieu equation,
\begin{equation}\label{eq:eom_quartic_mathieu}
	\frac{\diff^2 x_k}{\diff z^2} + \left( A_k - 2q\cos(2z) \right) x_k = 0 \, ,
\end{equation}
where we have defined
\begin{align}
	z &= \frac{3}{2} \sqrt{\lambda}\varphi_0 t_0 \left(a(t) - 1 \right) \pm \frac{\pi}{2} \, , \\
	A_k &= \left( \frac{4}{3} \right)^2 \frac{k^2}{\lambda \varphi_0^2} + 2q \, , \\ 
	q &= \frac{4}{9} \frac{g}{\lambda} \, .
\end{align}
Without the need to solve the differential equation for the mode functions explicitly, we know by Floquet's theorem that the solution contains an exponential factor,
\begin{equation}
	x_k \propto \exp \left( \mu_k z \right) \, ,
\end{equation}
with Floquet exponent $\mu_k$.
The Floquet exponents characterize the different solutions of the Mathieu equation.
For instance, if $\mu_k$ is imaginary, $x_k$ is an oscillatory function, falling into the class of stable solutions of the Mathieu equation.
We are however interested in the unstable solutions, i.e.~solutions with real and positive $\mu_k$, which correspond to exponential growth of the mode functions and thus rapid particle production.
The Floquet exponents are in turn characterized by so-called (in)stability bands of the Mathieu equation for different values of $A_k$ and $q$.
These stability bands determine the rate of resonant production of $\chi$-particles in the unstable regions, see, e.g., \cite{McLachlan:1951}.

The Floquet exponent $\mu_k$ for each mode is essentially set by $q$, i.e.~the ratio between the couplings $g$ and $\lambda$ of the theory.
This ratio is therefore the defining quantity that controls the production rate of $\chi$-particles.
Semi-analytic estimates for the Floquet exponent exist for two different limits, $q \ll 1$ and $q \gg 1$.
These two regimes are coined \textit{narrow} and \textit{broad} resonance, respectively.

\begin{figure}[t]
\centering
	\includegraphics[width=0.46\textwidth]{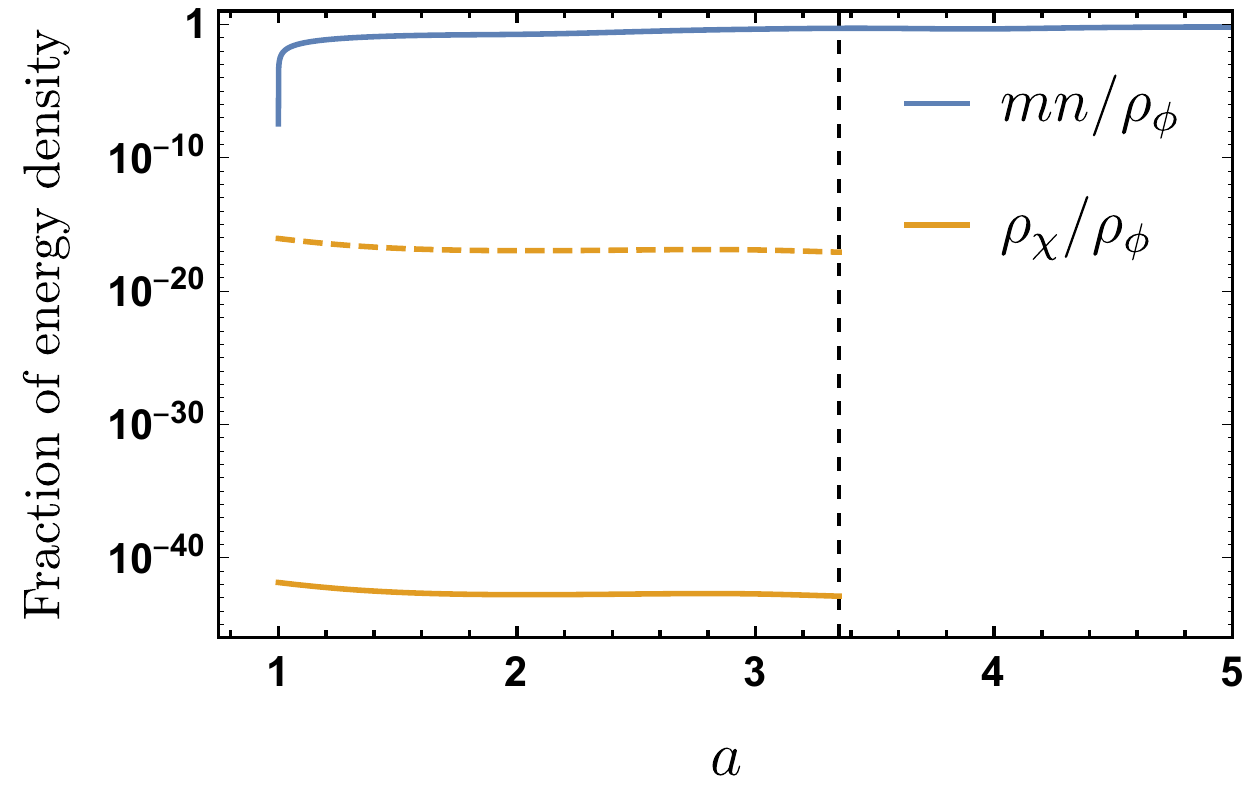}
	\hspace*{0.8cm}
	\includegraphics[width=0.46\textwidth]{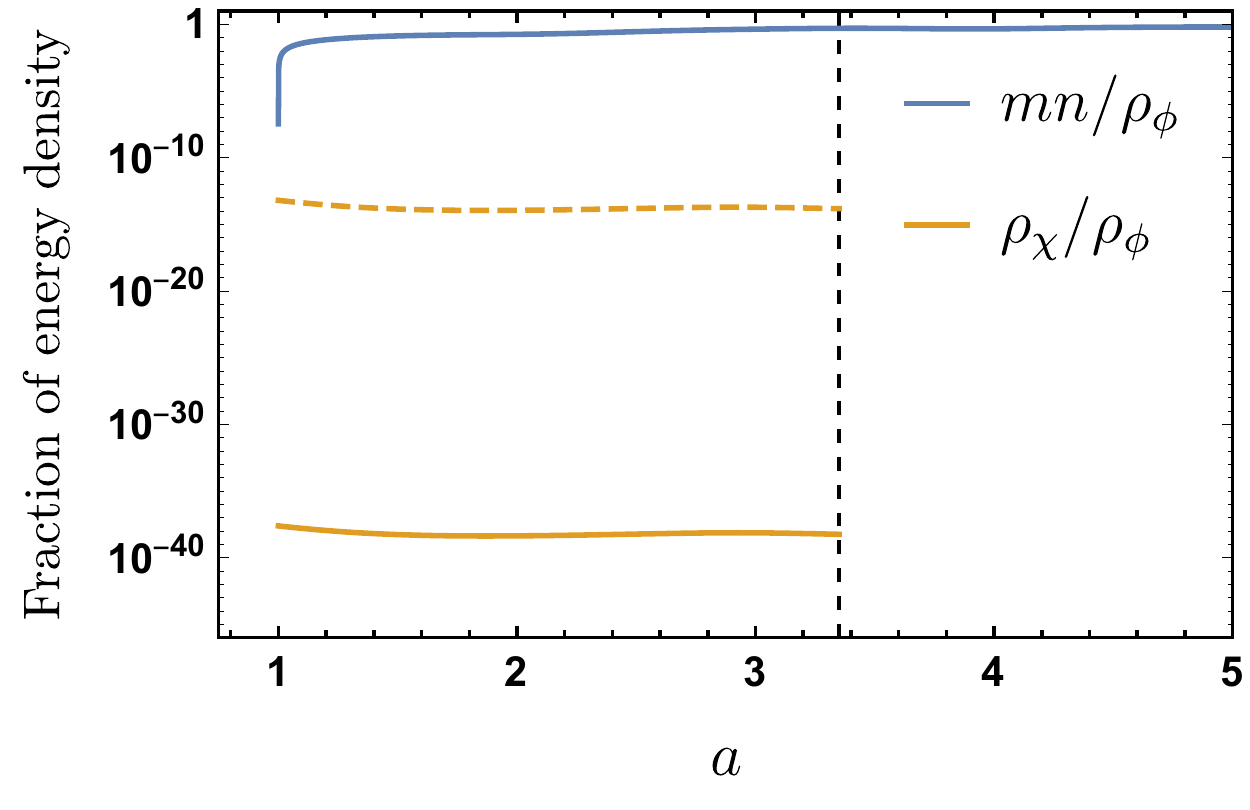}
	\caption{Fraction of energy density stored in charged particles compared to the fraction of the energy transferred to the field $\chi$ as a function of the scale factor $a$ for the quartic-dominated epoch. The end of the quartic-dominated regime is indicated by a dashed vertical line. As a benchmark for the coherent oscillations of $\phi$, we chose the parameters $m=1$ eV, $\lambda = 10^{-39}$, $\epsilon = 0.1\, \lambda$ and $\theta_0 = 0.2 \, \pi$. Furthermore, the initial field value is chosen to give the correct relic abundance of dark matter. The left panel corresponds to a narrow resonance, $g=0.1\, \lambda$, while the right panel shows a broad resonance, $g=10^4 \, \lambda$. The solid yellow line assumes that the initial occupation numbers are just the minimal quantum $n_{k}=1/2$, whereas the dashed line uses inflationary fluctuations $n_{k}\sim (H_{\rm I}/k)^2$, with $H_{\rm I}=10^4\,{\rm GeV}$. We note that our estimate accounts only for the exponential growth of the mode functions and does not take into account polynomial pre-factors. This causes the slight decrease in both panels.}
\label{fig:depletion_quartic}
\end{figure}

In Fig.~\ref{fig:depletion_quartic} we show an example of the energy transfer from the coherent oscillations of the symmetric part of $\phi$ to the field $\chi$ for both types of resonance as a function of the scale factor.
A detailed discussion of how to arrive at these results is given in Appendix~\ref{app:parametricresonance}.
The fraction of energy density stored in the created charged particles (shown in blue) is computed taking the full dynamics of the oscillations into account. However, the result for the resonant production of $\chi$-particles (shown in yellow) is only valid in a purely quartic-dominated regime for the radial oscillations.
For the chosen benchmark point, we observe that the generation of charge stops at $a \sim \mathcal{O}(10)$, before the creation of $\chi$-particles becomes significant.
In fact, our description of energy transfer is only applicable up to $a \sim \mathcal{O}(3)$, which is the point where the purely quartic-dominated regime ends and the mass term takes over.
That is, during the quartic-dominated epoch, the transfer of the energy to $\chi$ cannot reach a significant amount, such that the depletion of the uncharged component of the dark matter is neither efficient in narrow resonance, $g \ll \lambda$, nor in the broad resonance regime, $g \gg \lambda$.
Furthermore, this implies that the induced backreaction on the dynamics of $\phi$ due to the creation of $\chi$-particles\footnote{In particular, the resonant production of $\chi$-particles effectively contributes an additional friction term $\sim \Gamma \dot{\phi}$ to the equations of motion of $\phi$, where $\Gamma$ denotes the rate of particle production.} can be neglected as long as the quartic interaction terms dominate the evolution of $\phi$.
This assures that the mechanism of charge production described in the previous section is not affected.

\subsection{Mass-dominated epoch}

Once the mass term starts to dominate the evolution of the field, the $U(1)$ symmetry of the theory is restored and the generated charge density is (approximately) conserved.
However, similar to the quartic-dominated regime, the uncharged component of $\phi$ is still subject to annihilation into $\chi$-particles.

After a convenient field redefinition, $x_k = a^{3/2} \chi_k$, the equation of motion for the modes reads
\begin{equation}\label{eq:eom_chik_massdominated}
	\ddot{x}_k + \left( \frac{k^2}{a^2} + \frac{3}{4} H^2 + g \varphi^2 \right) x_k = 0 \, .
\end{equation}
In contrast to the quartic-dominated epoch, the explicit dependence on the Hubble scale of the expansion does not drop out\footnote{Note that in a universe dominated by matter instead of radiation, the term involving $H^2$ would disappear.}.

As pointed out in the beginning of this section, we assume that the evolution of the radial mode is entirely governed by $U(1)$ symmetric interactions, i.e.~it is given by Eq.~\eqref{eq:radial_sol_mass}.
Using this, the equation of motion is of quasi Mathieu form,
\begin{equation}\label{eq:eom_mass_mathieu}
	\frac{\diff^2 x_k}{\diff z^2} + \left( A_k - 2q\cos(2z) \right) x_k = 0 \, ,
\end{equation}
where we have defined
\begin{align}
	z &= mt \, , \\ 
	A_k &= \frac{k^2}{a^2m^2} + \frac{3}{4} \frac{H^2}{m^2} + 2q \, , \\ 
	q &= \frac{g\Phi^2}{4m^2} \, .
\end{align}
Here, $\Phi$ denotes the amplitude of the radial oscillations of the symmetric component.
The term involving the Hubble parameter appears as a shift of the momentum in each mode by $k^2 \to k^2 + 3/4 a^2H^2$.
As we will later see, this shift is small and can be neglected.

Eq.~\eqref{eq:eom_mass_mathieu} is not precisely a Mathieu equation, because the characteristic parameters $A_k$ and $q$ explicitly depend on time through the scale factor.
Depending on the coupling $g$, the system might start out in a broad resonance, but, as time evolves, $q$ continuously decreases until it eventually ends up in a narrow resonance.
In addition, $q$ might significantly vary within a few oscillations of the field, such that the analysis based on the static stability chart of the Mathieu equation is not applicable anymore.
Instead, the dominant process in this setting will be the so-called \textit{stochastic resonance}~\cite{Kofman:1997yn}.
In this regime, the occupation number $n_k$ typically increases, but may sometimes decrease instead.
The reason for this is that the frequency of the modes $\chi_k$ changes drastically with each oscillation of $\varphi$.
Consequently, the phases of $\chi_k$ are practically uncorrelated at the successive stages of particle production, mildly diminishing the resonant effect.
For a detailed discussion of the stochastic resonance phenomenon, we refer the reader to the seminal work~\cite{Kofman:1997yn}.
Here, we closely follow their derivation in order to estimate the rate of resonant production of $\chi$-particles during the mass-dominated epoch.

A necessary condition for the resonance to be active is given by $q^2 m \gtrsim H$~\cite{Kofman:1997yn}.
Given the typical scale of the Hubble parameter in the mass-dominated regime, this roughly translates into $\sqrt{g} \Phi \gtrsim m$.
This condition equivalently reads $q \lesssim 1/4$, which means that we expect particle production to stop in the narrow resonance regime.
The same condition, in turn, allows us to estimate the time scale $t_f$ or the number of oscillations $N_f$ after which the resonance terminates.

Rapid production of $\chi$-particles occurs whenever the adiabaticity condition for the mode frequencies of $\chi$ is violated,
\begin{equation}
\label{eq:resonancecond}
	\omega_k^2 \lesssim \frac{\diff \omega_k}{\diff t} \, ,
\end{equation}
with the frequency given by $\omega_k^2 = \frac{k^2}{a^2} + \frac{3}{4} H^2 + g \varphi^2$.
Due to the oscillations of the driving field, the resonance is only active for a short period of time in the vicinity of each zero crossing $\varphi(t) = 0$.
This period is typically of order inverse to the characteristic momenta involved in the resonance, $\Delta t \sim k_{\star}^{-1}$.
These, in turn, can be estimated by
\begin{equation}
	\frac{k^2}{a^2} \lesssim k_{\star}^2(t) \equiv \sqrt{g} m \Phi(t) \, .
\end{equation}
Since $k_{\star}^2 \gg a^2 H^2$, the momentum shift induced by the Hubble scale is completely negligible, as claimed above.

\begin{figure}[t]
\centering
	\includegraphics[width=0.5\textwidth]{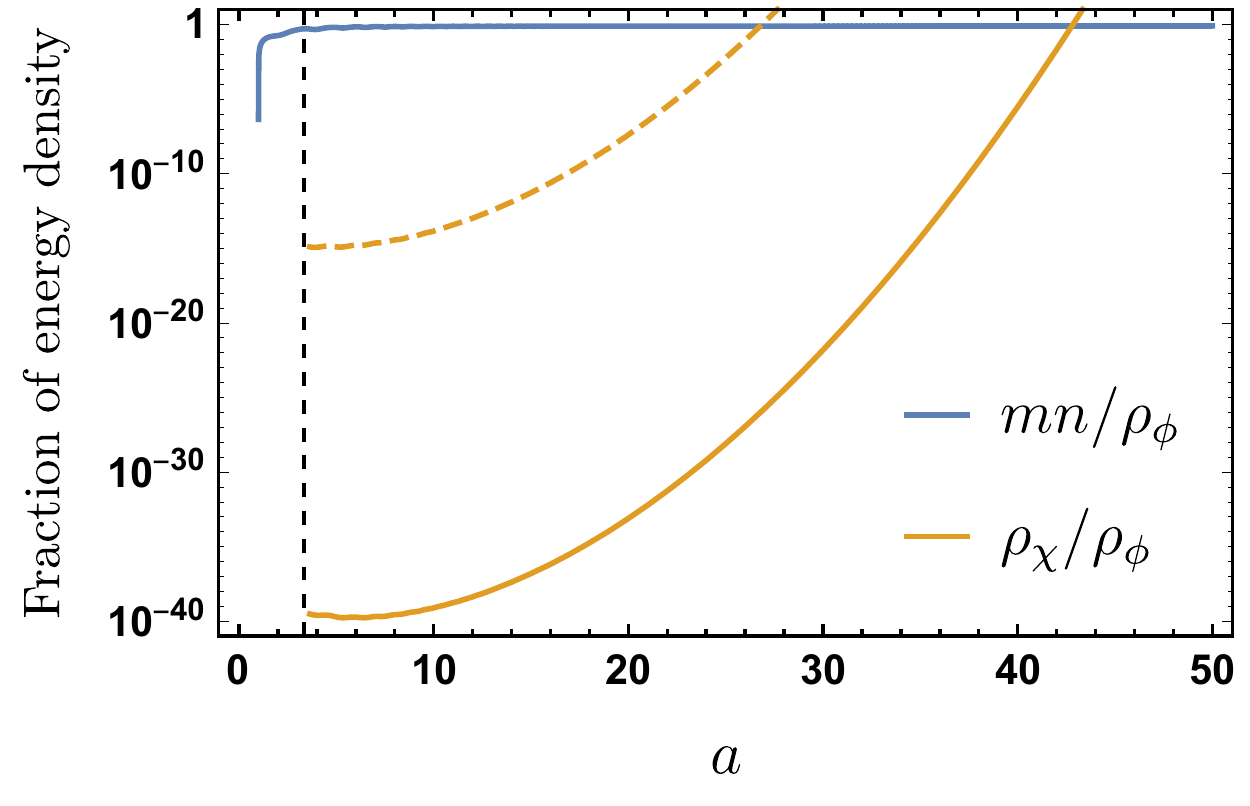}
	\caption{As in Fig.~\ref{fig:depletion_quartic}, fraction of energy density stored in charged particles compared to the fraction of the energy transferred to the field $\chi$ as a function of the scale factor $a$ but for the mass-dominated regime, which starts at the vertical dashed line. }
\label{fig:depletion_mass}
\end{figure}

Similar to what we did in the quartic-dominated epoch, we turn to the rate of the exponential growth of the occupation numbers,
\begin{equation}
n_k \sim \exp \left( 2 \mu_k m t \right) \, .
\end{equation}
That is, we want to estimate the Floquet exponent $\mu_k$ for the total duration of the resonance.
As worked out in detail in~\cite{Kofman:1997yn}, this exponent is given during each zero crossing $t_j$ by,
\begin{equation}\label{eq:floquetexp_mass}
	\mu_k^j = \frac{1}{2\pi} \ln \left( 1 + 2 e^{-\pi \kappa_j^2} - 2 \sin \hat{\theta} e^{-\frac{\pi}{2} \kappa_j^2} \sqrt{1 + e^{-\pi \kappa_j^2}} \right) \, ,
\end{equation}
with $\kappa_j^2 = k^2 / (a_j^2 \sqrt{g} m \Phi_j)$.
Here, each index labels an evaluation at $t_j$.
Furthermore, $\hat{\theta}$ denotes a random phase, implementing the fact that the phases of the modes are uncorrelated at successive moments of particle production.
For simplicity, we drop the term involving the random phase in Eq.~\eqref{eq:floquetexp_mass} for the remaining part of this work.

Each short burst of particle production contributes with a growth exponent $\mu_k^j \Delta t_j$.
In order to define an effective Floquet exponent, we average over the oscillations of $\varphi$,
\begin{equation}\label{eq:floquetexp_mass_avg}
	\mu_k = \frac{1}{t_f - t_i} \sum_{j=1}^{N_f} \mu_k^j \Delta t_j \, ,
\end{equation}
where $t_i$ and $t_f$ denote the start and the end of the resonance, respectively.
Using this averaged Floquet exponent~\eqref{eq:floquetexp_mass_avg}, the computation of the energy density that is transferred to the field $\chi$ proceeds as in the quartic-dominated regime (cf.~\eqref{eq:chi_energydensity}).

In Fig.~\ref{fig:depletion_mass} we illustrate an example of the energy transfer from the coherent oscillations of $\phi$ to the field $\chi$ as a function of the scale factor for a broad resonance, $g=10^4 \, \lambda$.
Similar to Fig.~\ref{fig:depletion_quartic}, the generation of charge density (shown in blue) takes the full dynamics into account, while for the resonant production of $\chi$-particles (shown in yellow) we assume a pure mass-dominated regime for the radial oscillations.
For the chosen benchmark point, in contrast to the quartic-dominated era, we observe that the energy transfer from $\phi$ to $\chi$ can be very efficient in the mass-dominated regime.

\bigskip

In the last few paragraphs of this section, we would like to point some relevant caveats to our estimations.
As it is, Fig.~\ref{fig:depletion_mass} seems to indicate that the symmetric part of the dark matter is entirely depleted before the resonance terminates, which for our particular choice of parameters is estimated to happen at $a_f\sim\mathcal{O}(70)$.
This picture is however not complete for two reasons.
Firstly, at the point where the energy transfer becomes an order one effect, our simplifications of neglecting backreaction break down and computational control is lost, preventing us from making precise statements about the exact amount of remaining symmetric abundance.
Secondly, we recall that so far we have only considered the $U(1)$ conserving terms for the evolution of $\phi$.
Including the asymmetric component (i.e.~the angular motion of the field) effectively introduces an additional mass term for $\chi$ that can act to either drive or even terminate the resonance, as we now discuss.

Indeed, taking into account the symmetric and asymmetric contributions to the radial mode (cf.~Section~\ref{subsec:charged})\footnote{Here, we drop the phase $\alpha$ which is irrelevant for our discussion.}, we have
\begin{equation}\label{eq:parametricresonance_asymmetric}
    \varphi^2 = \frac{\rho_{\rm CDM}}{m^2}\left[{\mathcal{A}}+{\mathcal{B}}\cos^{2}(mt)\right] \, ,
\end{equation}
where ${\mathcal{A}}$ grows with the asymmetric fraction of the energy density and ${\mathcal{B}}$ with the symmetric one.
The oscillating driving term is now proportional to ${\mathcal{B}}$.
In addition, we have a new constant part given by ${\mathcal{A}}$.
Inserting this into the condition~\eqref{eq:resonancecond} the numerator is $\sim {\mathcal{B}}$, whereas in the denominator we get a momentum shift by $k^2\to k^2+ g{\mathcal{A}}\rho_{\rm CDM}/m^2$.
The reason for this is that the field squared is always displaced from the minimum at least by an amount proportional to ${\mathcal{A}}$, and the effective mass of $\chi$ is therefore always positive.
Evaluating at the time when $\varphi^2$ is minimal, we find for the range of momenta that fulfill this condition,
\begin{equation}
    \left(\left(\frac{{\mathcal{B}}}{2}\right)^{\frac{2}{3}}c^{\frac{2}{3}}-{\mathcal{A}}c\right) \gtrsim \frac{k^2}{a^2 m^2}\gtrsim 1\, ,\quad{\rm where}\quad c=g\frac{\rho_{\rm CDM}}{m^2}\sim\frac{1}{a^3} \, .
\end{equation}
Here, the condition on the right hand side of $k^2$ corresponds to the need for the instability to be active over a sufficient momentum range.
Since ${\mathcal{B}}\lesssim 2$, the left hand side requires $c\gtrsim 1$ for the resonance to be active.
Looking now at the second term on the left hand side we see that the resonance is hindered by the asymmetric part.
When ${\mathcal{A}}\sim 1$ the resonance stops for all values of $c\gtrsim 1$. 
Using Eq.~\eqref{eq:ab} we find that this happens when $\eta_{\rm a}\sim \eta_{\rm s}$.
In practice, this mechanism therefore does not allow to effectively deplete the symmetric part far beyond the halfway point.

\bigskip

In summary, we conclude that for suitable values of the portal coupling, the energy transfer can be negligible during the quartic-dominated regime, but very efficient during the mass-dominated epoch.
This means that while the generation of the charge density is not affected, part of the energy density stored in the radial oscillations can be efficiently depleted, at least up to reaching the level where $\eta_{\rm a}\sim \eta_{\rm s}$.
Another interesting possibility would be to consider a negative portal coupling $g$, which can induce a tachyonic mass term for the $\chi$ field.
Such a tachyonic instability allows for the annihilation channel to be always kinematically open, in principle enabling a much more efficient depletion of the symmetric dark matter component.
As a matter of fact, this case can be treated with the same methods of parametric resonance that we presented in this section.
In line with~\cite{Greene:1997ge}, we find indications that the energy transfer from $\phi$ to $\chi$ can be tremendously increased already in the quartic-dominated epoch, as the resonance bands are much broader and the corresponding Floquet exponents are orders of magnitude larger, even for moderate values of the coupling.
The details of this interesting possibility for an efficient depletion of the energy density stored in the uncharged component of $\phi$ would require a dedicated study.

\section{Phenomenology of light dark matter with a net charge}
\label{sec:phenomenology}

As already outlined in the introduction, detection of very light dark matter carrying a non-vanishing charge density is made more challenging mainly by two effects.

The first one is that only the combination $\phi^{\dagger}\phi$ can appear in couplings, instead of just $\phi_{\rm real}$ as would be the case for a real scalar field $\phi_{\rm real}$.
This severely limits the available interactions between the dark matter and the Standard Model.
Indeed, at the renormalizable level there is only one ``portal'' coupling
\begin{equation}\label{eq:phihcoupling}
	\mathcal{L}_H = \kappa \abs{\phi}^2 H^\dagger H \, 
\end{equation}
which is the one that will be considered in more detail in the next subsection.

The challenges that arise due to the $U(1)$ symmetry are perhaps more obvious for higher dimensional operators. For example, the usual two-photon coupling of an axion-like particle is replaced by\footnote{A CP even scalar would couple to $F^2$, while allowing for CP violation the coupling to $F\tilde{F}$ could be the relevant one. This does not affect our dimensional argument.}
\begin{equation}
    \frac{1}{M}\phi_{\rm real} F\tilde{F} \quad \hookrightarrow \quad \frac{1}{M^2} \abs{\phi}^2 F^2.
\end{equation}
We see that higher dimensional operators generically feature an extra suppression by the potentially large energy scale $M$.

\bigskip

The second challenge, which is closely related to the first one, is that charged particles cannot be absorbed, an effect that is used in most experiments for the detection of very light dark matter particles. 

Let us have a more careful look at this effect. Very light dark matter particles have very high occupation numbers and can be conveniently described by using classical field equations (as we have done in this paper). 
For a real scalar field and neglecting velocity effects\footnote{This is a good approximation within the coherence length $\sim 1/(m v)$, since the velocity is very non-relativistic.}, the effect of dark matter in a typical experimental setup corresponds to a spatially constant oscillating field,
\begin{equation}
    \phi_{\rm real}\sim \Phi_{\rm real}\sin(mt) \, .
\end{equation}
Considering typical interactions with the Higgs boson or photons we have
\begin{equation}
    b \, \Phi_{\rm real}\sin(mt)H^{\dagger}H \quad {\rm or} \quad \frac{\Phi_{\rm real}}{M}\sin(mt)F\tilde{F} \, .
\end{equation}
where $b$ denotes the coupling constant of the real Higgs portal that has mass dimension~1.
In other words, an oscillating term appears in front of the operators involving Standard Model fields.
This implies that, from the point of view of the visible sector, time translation invariance is broken and the Standard Model fields can gain energy.
Under suitable conditions, the absorption of unconserved dark matter particles can take the form of oscillating driving forces.
For example, the Higgs interaction leads to oscillating mechanical forces on test masses~\cite{Graham:2015ifn} whereas the photon interaction can drive an electromagnetic resonator~\cite{Sikivie:1983ip}.

Let us now see how the situation changes for the case of charged scalar fields.
As discussed in Section~\ref{subsec:charged} the charge conserving interaction term can be readily expressed in terms of the density as in Eq.~\eqref{eq:invariant}.
Let us consider the case, when most of the energy density is stored in the asymmetric component, i.e. $1-\eta_{\rm a}=\eta_{\rm s}\ll 1$.
In this case Eq.~\eqref{eq:invariant} reads\footnote{We again drop the irrelevant phase $\alpha$.},
\begin{equation}
    \phi^{\dagger}\phi=\frac{1}{2}\frac{\rho_{\rm CDM}}{m^2}\left[1+2\sqrt{2\eta_{\rm s}\eta_{\rm a}}\cos^{2}(mt)\right].
\end{equation}
If there is no symmetric part, $\eta_{\rm s}=0$, there is no oscillating term and the Standard Model fields cannot gain energy in this way.
This can be understood because any absorption can only proceed as a pairwise particle-antiparticle annihilation, and if there is only one species, this cannot happen.
Indeed, the oscillating terms are proportional to $\sqrt{\eta_{\rm s}}$, showing that an uncharged, symmetric component is needed.
The power absorbed is typically proportional to the square of the oscillating term and therefore $\sim \eta_{\rm a}\eta_{\rm s}$.
This can be roughly understood as the probability of a particle meeting a suitable antiparticle at each interaction.

\subsection{The Higgs portal}

Let us briefly illustrate our general considerations above for the concrete example of the only renormalizable portal for a charged dark matter particle, i.e.~the Higgs portal coupling given in Eq.~\eqref{eq:phihcoupling}.

\subsubsection*{Fine tuning}
The Higgs portal coupling induces a contribution to the mass of $\phi$,
\begin{equation}
	m^2 = m_0^2 +  \kappa v_0^2/2 \, ,
\end{equation}
where $v_0 = 246$~GeV is the measured value of the Higgs vacuum expectation value (vev).
If the field $\phi$ is to remain light and without invoking a precise cancellation between the two different contributions to the mass, we require that
\begin{equation}
	\kappa \lesssim 3\cdot 10^{-35}\left(\frac{m}{\mu\text{eV}}\right)^2 \, .
\end{equation}
This should however be seen as a purely fine tuning argument rather than a strict limit\footnote{The Higgs vev is in general a dynamical quantity that varies during the history of the Universe. In order not to experience a dramatic change in the cosmological evolution of $\phi$ (or avoid a dynamical fine tuning), we can assume that the reheating temperature never exceeds the electroweak scale so that the Higgs vev is fixed to its current value during the whole postinflationary epoch.}.
In the following, we want to turn to direct experimental tests and observations.

\subsubsection*{Higgs decays}
If the dark matter is much lighter than the Higgs, $m\ll m_H/2$, the current experimental constraint on the branching ratio ${\rm BR} (H \to \rm{inv.}) < 0.24$~\cite{Tanabashi:2018oca} bounds $\kappa$ to be 
\begin{equation}
    \kappa\lesssim 0.0036 \, ,
\end{equation}
which is in line with, e.g.,~\cite{Arcadi:2019lka}. This is a relatively weak constraint, but importantly does not rely on $\phi$ being dark matter.

\subsubsection*{Varying fermion masses}
The Higgs portal coupling~\eqref{eq:phihcoupling} also leads to a shift of the mass parameter $\mu^2$ of the Higgs potential or, equivalently, to a shift of the Higgs vev,
\begin{equation}
	v = \sqrt{\frac{\mu^2-\kappa \abs{\phi}^2}{\lambda}} \approx v_0 \left( 1 - \frac{\kappa \abs{\phi}^2}{2\mu^2} \right) \, ,
\end{equation}
where we define $v_0=\sqrt{\mu^2 / \lambda}$, which corresponds to the unperturbed ($\kappa=0$) Higgs vev.
The latter expression assumes that the shift is small, and is valid as long as $ \kappa \abs{\phi}^2 \ll \mu^2$.
Clearly, a modified Higgs vev translates into a shift of the Standard Model fermion masses,
\begin{equation}
	m_f \approx m_f^0 \left( 1 - \frac{\kappa \abs{\phi}^2}{2\mu^2} \right).
\end{equation}
The correction to the mass is hence $\Delta m_f=m_f^0 \kappa \abs{\phi}^2 /(2 \mu^2)$, which can be positive or negative depending on the sign of the coupling $\kappa$.

Since in our setup we assume that the totality of the dark matter is explained by the scalar field, the relative modification of the fermion masses can be rewritten in terms of the dark matter density,
\begin{equation}
	\frac{\Delta m_f}{m_f^0} =  \frac{\kappa}{4\mu^2} \frac{\rho_{\mathrm{DM}}}{m^2} \, .
\end{equation}
In this form, the dependence of the correction on the mass of $\phi$ is also made explicit.
Parametrically, the relative change in the fermion masses is
\begin{equation}\label{eq:fermion_mass_shift}
	\frac{\Delta m_f}{m_f^0} \simeq  10^{-16}\kappa \left( \frac{\rho_{\mathrm{DM}}^{\rm local}}{0.4~\text{GeV}/\text{cm}^3
	}\right)\left(\frac{\mu\text{eV}}{m}\right)^2 \, ,
\end{equation}
where we have used the measured value for the Higgs mass $m_H\simeq 125$~GeV~\cite{Aad:2012tfa,Chatrchyan:2012xdj,Tanabashi:2018oca}.
An absolute shift in the fermion masses is of course hard to observe, but the fact that Eq.~\eqref{eq:fermion_mass_shift} depends on the (local) dark matter density opens up interesting possibilities.
The change in the fermion mass can for example affect the frequency of atomic transitions.
Current limits coming from atomic clocks bound the time variation of the ratio $\mu=m_p/m_e$ to be~\cite{Godun:2014naa}
\begin{equation}
\frac{\dot{\mu}}{\mu}=(0.2\pm 1.1)\cdot10^{-16}\text{yr}^{-1}~.
\end{equation}
Similar bounds have also been derived in \cite{Huntemann:2014dya}.
In our setup, variations of $\mu$ can arise if the dark matter density at the experimental locations (i.e.~the Earth) changes over time due to, for instance, the presence of very small scale dark matter substructure.
Assuming an observation time of 1 year, $\kappa$ could be constraint to be
\begin{equation}
\kappa\lesssim 2\cdot10^{-1} \left( \frac{0.4~\text{GeV}/\text{cm}^3
	}{\rho_{\mathrm{DM}}^{\rm local}}\right)\left(\frac{m}{\mu\text{eV}}\right)^2\left(\frac{\Delta\rho_{\rm DM}}{\rho_{\rm DM}^{\rm local}}\right)^{-1}~,
\end{equation}
where $\Delta\rho_{\rm DM}$ denotes the typical change in the local dark matter density over the appropriate observation time.

\subsubsection*{Big Bang nucleosynthesis}
Apart from generating a shift in the fermion masses\footnote{In spirit, the limit derived here from BBN is similar to the one obtained in~\cite{Stadnik:2015kia,Stadnik:2015uka} for the very low mass limit $m\ll 10^{-16}\,{\rm eV}$ of a real scalar quadratically coupled to fermions. There, the field value is not oscillating during BBN either.}, the modification of the Higgs vev also affects the Fermi constant,
\begin{equation}\label{eq:shift_fermi_constant}
	G_F = \frac{1}{\sqrt{2} v^2} = G_F^0 \left(1 -  \frac{\kappa}{4\mu^2} \frac{\rho_{\mathrm{DM}}}{m^2} \right)^{-2} \, .
\end{equation}
Hence, the DM-Higgs coupling alters all weak interaction rates.
A powerful probe of weak interactions and the Fermi constant in the early universe around $T \approx 1$ MeV is provided by Big Bang nucleosynthesis (BBN).
Detailed studies of the effect of a shift in the Fermi constant on BBN have been conducted in, e.g.,~\cite{Kolb:1985sj,Scherrer:1992na}.
Here, we quote their results and apply them to our setup.
The observed abundance of $^4$He sets lower and upper bounds on the ratio of Fermi constant at BBN compared to the value measured today,
\begin{equation}
	0.78 < G_F^{\rm BBN}/G_F^0 < 1.01 \, .
\end{equation}
The lower bound increases to 0.85 when the limits on $^7$Li production are also included. We can translate the constraints into a bound on $\kappa$ by taking into account the redshift of the dark matter density\footnote{For $m\lesssim10^{-16}$~eV, the field does not start rolling until after BBN and the redshift dependence assumed here is not directly applicable~\cite{Blum:2014vsa}.}. This gives
\begin{align}
-4\cdot 10^{-9}\lesssim\kappa \left(\frac{\mu\text{eV}}{m}\right)^2\left(\frac{\rho_{\text{DM}}^{\rm av}}{1.3~\text{keV}/\text{cm}^3}\right)\lesssim 2\cdot 10^{-10}\,.
\end{align}
The lower bound changes to $-3\cdot 10^{-9}$ if the effect on $^7$Li production is taken into account.	
Despite the measurement itself being way less precise, this constraint is much stronger than the one coming from atomic clocks.
The reason for this is that the dark matter density at the time of BBN was about $20$ orders of magnitude larger than the present (local) one, and therefore the shift in Eq.~\eqref{eq:shift_fermi_constant} benefits from this huge enhancement.

\subsubsection*{Astrophysical observations}
Some astrophysical observables can also be used to constrain variations in $G_F$.
For example, the agreement between the redshift determined from the 21~cm line and the redshift determined from optical resonance lines of the same object constrains the electron-to-proton mass ratio $m_e/m_p$, the fine structure constant $\alpha$ as well as the product $\alpha^2 g_p m_e/m_p$~\cite{Bahcall:1967zz,Wolfe:1976zz}, where $g_p$ denotes the gyromagnetic moment of the proton.
Since the bulk of $m_p$ does not come from the Higgs mechanism, its relative change in the proton mass is negligible compared to the electron one.
The strongest limits come from a quasar absorption system at $z=1.77$, which constrain the product $\Delta \ln \left( \alpha^2 g_p m_e/m_p \right) < 2 \times 10^{-4}$~\cite{Wolfe:1979}.
This implies that $G_F$ varies less than $0.04\%$ from today to that redshift.
Furthermore, from the agreement between optical and radio redshifts of a distant galaxy at $z=3.4$, it has been determined that $G_F$ varies less than $0.2\%$ back to a redshift of $z=3.4$~\cite{Uson:1991zz}.
Translating these constraints into bounds on $\kappa$ gives
\begin{align}
\abs{\kappa} &\lesssim 
2\cdot10^{12} \left(\frac{m}{\mu\text{eV}}\right)^2\left(\frac{0.4 ~\text{GeV}/\text{cm}^3}{\rho_{\text{DM}}^{\rm local}}\right)\left(\frac{\Delta\rho_{\rm DM}}{\rho_{\rm DM}^{\rm local}}\right)^{-1}
,~\text{at }z=1.77 \, , \\
\abs{\kappa} &\lesssim 
1\cdot10^{13} \left(\frac{m}{\mu\text{eV}}\right)^2\left(\frac{0.4 ~\text{GeV}/\text{cm}^3}{\rho_{\text{DM}}^{\rm local}}\right)\left(\frac{\Delta\rho_{\rm DM}}{\rho_{\rm DM}^{\rm local}}\right)^{-1},~\text{at }z=3.4 \, .
\end{align}
This result depends on the difference between the dark matter density of the observed galaxy to the local one at Earth, which is denoted by $\Delta\rho _{\rm DM}$.

\section{Conclusions}
\label{sec:conclusion}

In this work, we have studied a model of very light (even sub-eV) asymmetric dark matter.
In our setup, dark matter is produced non-thermally via the Affleck-Dine mechanism.
The final energy density is a combination of a symmetric component, i.e.~one that contains an equal number of particles and antiparticles, and an asymmetric one only constituted of particles (or antiparticles).
These two components can naturally be of the same size if the symmetry breaking terms in the potential are comparable to the symmetric self-interactions, and for particular initial conditions the asymmetric component can contribute in excess of $\sim 95\%$ to the total dark matter abundance.

We have also described how a coupling $\abs{\phi}^2\chi^2$ of the charged field $\phi$ to an additional (massless) scalar field $\chi$ with a positive coupling constant allows to deplete the symmetric part of the energy density via a parametric resonance effect.
The symmetric component cannot completely disappear in this way, as this mechanism becomes inefficient when the two components are of the same size.
Therefore, if such couplings are present, the present-day dark matter abundance is generically expected to be evenly distributed between a symmetric and an asymmetric population.
A negative coupling constant leading to a tachyonic instability would make the depletion more efficient, but also cause a breakdown of the approximations on which our results rely.
We leave a more complete study of this situation to future work.
A promising way forward could be an approach, e.g., based on classical statistical field theory simulations.
Such an approach would also automatically include the effects of fragmentation, i.e.~the disintegration of the homogeneous field into inhomogeneous fluctuations.
This effect generally occurs in a large part of parameter space where the self-interactions are somewhat larger than the ones we study here (which correspond to the minimal values required to efficiently produce a charge asymmetry).
Sufficiently large self-interactions could also have other consequences such as the formation of Q-balls or other macroscopic, inhomogeneous configurations that may have interesting effects for phenomenology.
These possibilities merit further investigation.

A crucial feature of such light asymmetric dark matter is the change in the resulting phenomenology and, in particular, of the expected experimental signatures.
Direct detection of sub-eV dark matter usually proceeds via an effective absorption of the dark matter particles.
The existence of a comovingly conserved particle number in the dark matter however forbids the asymmetric component from being absorbed, making this detection strategy much less sensitive.
Further suppression arises from the restrictions on the interaction with Standard Model particles due to the requirement of charge conservation.

As an application of this idea, we study the simple example of a Higgs portal coupling interaction.
Instead of the usual oscillatory time variations of constants of nature (such as fermion masses or $G_F$), in the asymmetric scenario effects only arise due to variations in the dark matter density, therefore hindering the sensitivity of the traditional experimental setups.
To overcome this limitation, we study alternative probes based on variations arising from redshift evolution as well as local fluctuations of the dark matter field.

It is of course also possible to look for the symmetric component of the dark matter, which generically appears in our model but can be comparatively subdominant to the asymmetric one.
In any case, this symmetric component is susceptible to the usual resonant search approaches looking for oscillations at a fixed frequency\footnote{More precisely, this would look similar to the real scalar case with quadratic fermion couplings studied, e.g., in~\cite{Stadnik:2015kia,Stadnik:2015uka}.}, as the ones that are employed to look for dark matter made from light uncharged bosons.
Similar effects would also arise from suppressed charge violating couplings to the Standard Model. 

Finally, we would like to point out that, since the sensitivity of direct detection searches is reduced for very light asymmetric dark matter, its couplings to the visible sector are allowed to be larger than is commonly assumed in the usual uncharged scenario.
This opens up the opportunity to search for such particles in laboratory or astrophysical environments that can directly produce them.

\section*{Acknowledgements}

We are grateful to Doddy Marsh for useful discussions on structure formation. J.~G.~thanks Heidelberg University for kind hospitality in the early stages of the project. G.~A.~receives support from the Fundaci\'on ``la Caixa'' via a ``la Caixa" postgraduate fellowship.
S.~S.~gratefully acknowledges financial support by the Heidelberg Graduate School for Physics during which major parts of this work were carried out.
This project has received support from the European Union's Horizon 2020 research and innovation programme under the Marie Sklodowska-Curie grant agreement No 674896 and No 690575.

\appendix

\section{Parametric resonance and the Mathieu equation}
\label{app:parametricresonance}

The resonant production of particles in an oscillating background field is described by the Mathieu equation,
\begin{equation}\label{eq:mathieu}
	\frac{\diff^2 x_k}{\diff z^2} + \left( A_k - 2q\cos(2z) \right) x_k = 0 \, .
\end{equation}
Floquet's theorem implies that the solutions to the Mathieu equation contain an exponential factor,
\begin{equation}
	x_k \sim \exp \left( \mu_k z \right) \, .
\end{equation}
The Floquet exponents $\mu_k$ are characterized by so called (in)stability bands of the Mathieu equation for different values of $A_k$ and $q$, see, e.g.,~\cite{McLachlan:1951}.

We want to discuss two different regimes, $q \ll 1$ and $q \gg 1$, i.e.~the \textit{narrow} and \textit{broad} resonance.
Both are important to determine the rate of resonant $\chi$-particle production in the classical background field $\phi$.
The equations of motion for the mode functions $x_k$ are given in Section~\ref{sec:depletion}.

\subsection{Narrow resonance}

In the regime of narrow resonance, $q \ll 1$, the dominant contribution to the unstable solution comes from the first resonance band\footnote{See, for instance, the stability chart of the Mathieu equation in~\cite{McLachlan:1951}.} of the Mathieu equation.
In this band, the Floquet exponent is given by~\cite{McLachlan:1951,Kofman:1997yn}
\begin{equation}
	\mu_k = \sqrt{\left( \frac{q}{2} \right)^2 - \left( \frac{k}{\Lambda} - 1 \right)^2} \, ,
\end{equation}
where we defined $\Lambda^2 \equiv (3/4)^2 \lambda \varphi_0^2$.
It is maximal at its center, $\mu_{k_\star} = q/2$, which corresponds to the momentum $k_{\star}^2 \sim \lambda \varphi_0^2$.
In this sense, the resonance is narrow, because only very few, distinct momentum modes are enhanced -- in this case it is the specific mode $k_{\star}$.
Integrating over the narrow momentum band (cf.~Eq.~\eqref{eq:chi_energydensity}) yields the transfer of energy from $\phi$ to $\chi$.
It occurs at a rate
\begin{equation}
	\Gamma \sim \frac{2}{3} \frac{g \varphi_0}{\sqrt{\lambda}} \frac{1}{a(t)} \, .
\end{equation}

Note that this rate is larger than the naive expectation from a perturbative approach.
In particular, treating the interaction term~\eqref{eq:interaction} as an effective (time dependent) trilinear coupling $g \langle \varphi \rangle \varphi \chi^2$, one can perturbatively study the decay $\varphi \to \chi \chi$.
Parametrically, the corresponding rate is given by $\Gamma \sim \frac{g^2 \varphi_0}{\sqrt{\lambda}} \frac{1}{a(t)}$, which is suppressed by an additional factor of the coupling compared to the parametric resonance.
On the one hand, this additional factor can be recovered by also taking Bose enhancement into account, see, e.g., \cite{Baumann}.
On the other hand, from the perspective of parametric resonance one can in principle also recover the perturbative result, when considering the (properly normalized) energy transfer from $\phi$ to $\chi$ at early times.
In this sense, both methods describe the same outcome from different perspectives.

\subsection{Broad resonance}

The broad resonance regime corresponds to $q \gg 1$.
As its name suggests, in contrast to the narrow resonance, it corresponds to a regime where a broader spectrum of modes is enhanced.
Therefore, particle production is typically more efficient compared to a narrow resonance.

The dynamics of the Mathieu equation are less under control in the broad resonance case compared to the narrow resonance regime.
Still, especially if $q \gg 1$, we can make use of an analytic approximation of the Floquet exponents given in~\cite{Fujisaki:1995ua},
\begin{equation}
	\mu_k = \frac{1}{\pi} \log \left( \sqrt{x} + \sqrt{x-1} \right) \, ,
\end{equation}
where we define
\begin{align}
    x &= \left( 1+ e^{-\pi\sqrt{q}\epsilon} \right) \cos^2\Psi \, ,\\
    \Psi &= \left( \frac{\pi^2}{2} + \epsilon\log\left( \pi q^{1/4} \right) \right) \sqrt{q} + \frac{1}{2} \Im \left[ \frac{\Gamma\left( \left( 1 - i\sqrt{q}\epsilon\right) / 2 \right)}{\Gamma\left( \left( 1 + i\sqrt{q}\epsilon\right) / 2 \right)} \right] \, .
\end{align}
In terms of the parameters of the Mathieu equation, we furthermore define $\epsilon\equiv\frac{A_k}{2q}-1$.
Here, we consider the case $\epsilon > 0$.
For the regime $\epsilon < 0$, see~\cite{Fujisaki:1995dy}.

Using the analytic approximation of $\mu_k$, it is straightforward to compute the energy density stored in the creation of particles by numerically evaluating the momentum integral given in Eq.~\eqref{eq:chi_energydensity}.

\end{document}